\documentclass[prb, twocolumn, showpacs, amsmath, amssymb,superscriptaddress]{revtex4-1}

\usepackage{amssymb,amsfonts,amsmath} 
\usepackage{graphicx}
\usepackage{color}
\usepackage{hyperref}
\usepackage{longtable}
\usepackage{bbm}
\usepackage{ulem}
\begin{document} 

\title{Small quenches and thermalization}

\author{D.M.\ Kennes}
\affiliation{Institut f{\"u}r Theorie der Statistischen Physik, RWTH Aachen University 
and JARA---Fundamentals of Future Information
Technology, 52056 Aachen, Germany}
\affiliation{Department of Physics, Columbia University, New York, NY 10027, USA}
\author{J.C.\ Pommerening} 
\affiliation{Institut f{\"u}r Theorie der Statistischen Physik, RWTH Aachen University 
and JARA---Fundamentals of Future Information
Technology, 52056 Aachen, Germany}
\author{J.\ Diekmann}  
\affiliation{Institut f{\"u}r Theorie der Statistischen Physik, RWTH Aachen University 
and JARA---Fundamentals of Future Information
Technology, 52056 Aachen, Germany}
\author{C.\ Karrasch} 
\affiliation{Dahlem Center for Complex Quantum Systems and Fachbereich Physik, 
Freie Universit\"at Berlin, 14195 Berlin, Germany}
\author{V.\ Meden} 
\affiliation{Institut f{\"u}r Theorie der Statistischen Physik, RWTH Aachen University 
and JARA---Fundamentals of Future Information
Technology, 52056 Aachen, Germany}

\begin{abstract} 
We study the expectation values of observables and correlation functions at long times 
after a global quantum quench. Our focus is on metallic (`gapless') 
fermionic many-body models and small quenches.  The system is prepared in an 
eigenstate of an initial Hamiltonian, and the time evolution is performed with a 
final Hamiltonian which differs from the initial one in the value of one global 
parameter. We first derive general relations between time-averaged expectation values 
of observables as well as correlation functions and those obtained in an eigenstate of 
the final Hamiltonian. Our results are valid to linear and quadratic order in the quench 
parameter $g$ and generalize prior insights in several essential ways. This allows us to 
develop a phenomenology for the thermalization of {\it local} quantities up to a given order 
in $g$. Our phenomenology is put to a test in several case studies of one-dimensional models 
representative of four distinct classes of Hamiltonians: quadratic ones, effectively quadratic 
ones, those characterized by an extensive set of (quasi-) local integrals of motion, and 
those for which no such set is known (and believed to be nonexistent). We show that for 
each of these models, all observables and correlation functions thermalize to linear 
order in $g$. The more {\it local} a given quantity, the longer the linear behavior prevails 
when increasing $g$. Typical {\it local} correlation functions and observables for 
which the term ${\mathcal O}(g)$ vanishes thermalize even to order $g^2$.
Our results show that lowest order thermalization of 
{\it local} observables is an ubiquitous phenomenon even in models with extensive sets of 
integrals of motion.  
\end{abstract}

\date{\today} 
\maketitle

\section{Introduction}
\label{sec:intro}

\subsection{Relaxation and thermalization}

Over the past decade, the question if and how expectation values of observables as well as 
correlation functions of 
closed quantum many-body systems time evolved with a (final) Hamiltonian $H_{\rm f}$ 
approach a steady-state value was heavily investigated.\cite{Polkovnikov11,Gogolin15} 
Nontrivial relaxation in the large-time limit $t \to \infty$ can only occur if the initial state 
is characterized by a density matrix $\rho_{\rm i}$ which does not commute with $H_{\rm f}$.  
If at least a few quantities reach steady-state values, the obvious question arises whether or not these can consistently be computed via a time-independent statistical operator, and if this operator can be chosen as the density matrix of one of the standard ensembles of 
equilibrium statistical mechanics. The latter scenario is widely known as {\it thermalization}. 

While there are well-established concepts to answer these questions under rather 
general conditions in classical mechanics,\cite{Gogolin15}  the situation is by far less clear -- and theoretically more challenging -- in the 
quantum case. Given the recent experimental advances in preparing and controlling cold atomic gases, it is now possible to realize quantum 
many-body systems which can be viewed as isolated.\cite{Bloch08} 
Gaining a comprehensive understanding of the nonequilibrium dynamics of such systems is 
thus a pressing problem of theoretical quantum physics. 

\subsection{Quench dynamics}

Answering the above questions in full generality is still out of reach. Hence, prior works usually focused on investigating certain special scenarios. One of these is the dynamics (and possible steady state) resulting 
out of a global quantum quench. In this protocol, one assumes that the system is prepared 
in an equilibrium state of an initial Hamiltonian $H_{\rm i}$ (e.g., the ground state 
$\left|E^{\rm i}_0\right>$) and that it is subsequently time evolved with $H_{\rm f}$. 
Both Hamiltonians differ by the value of at least one global parameter $g$, which is often taken to be the two-particle interaction of the many-body Hamiltonian, defining the subclass of `interaction quenches'. Quantum quenches can be realized experimentally 
in cold gases since those setups allow one to change a global parameter on time scales that are short 
compared to the internal ones.\cite{Bloch08} 

Only a few model-independent results on the quench dynamics are 
available.\cite{Polkovnikov11,Gogolin15,DAlessio15}
Most studies of concrete models focused on one-dimensional (1d) systems for which a variety of analytical as well as numerical nonequilibrium many-body methods exist to tackle the dynamics. However, even the quench problem in 1d is still far from being understood 
completely. 

In this paper, we exclusively focus on the case of {\it small quenches} 
($g\ll1$).\cite{Moeckel08,Moeckel09,Calabrese11,Calabrese12} 
This allows us to gain unbiased insights, and we do not need to make use of advanced 
ideas such as `quantum integrability'\cite{Gogolin15} or the eigenstate thermalization 
hypothesis.\cite{DAlessio15} We generalize the analytical, model-independent results of Refs.~\onlinecite{Moeckel08} and  \onlinecite{Moeckel09} and propose a `phenomenological picture of small quenches'. We then test this phenomenology explicitly for a variety of 1d models, which (depending on their complexity) we solve analytically or numerically using matrix product state techniques.\cite{Schollwoeck11} In particular, we exploit 
the fact that the latter can be implemented directly (and elegantly) in the thermodynamic limit to 
obtain results which are free of finite-size effects.\cite{Karrasch12a}

\subsection{Time averages for small quenches}
\label{smallquenchintro}

Considering quenches in which the amplitude $g$ is small offers the perspective
of obtaining model-independent, analytical results using perturbation theory. This was 
exploited in Refs.~\onlinecite{Moeckel08} and  \onlinecite{Moeckel09}, whose results we now briefly summarize.

{\it Discussion of prior results ---} In the above-mentioned works, it was shown that for a
finite system the long-time limit of the time-averaged expectation value
\begin{eqnarray}
\overline{\left< O \right>} = \lim_{\tau \to \infty} \frac{1}{\tau} \int_0^\tau 
\left< E_0^{\rm i} \right| O(t) \left| E_0^{\rm i} \right>dt ,
\label{blarichtig}
\end{eqnarray}  
with $O(t)$ being the observable in the Heisenberg picture with respect to $H_{\rm f}$, equals two times the equilibrium expectation value of 
$O$ in the ground state $ \left| E_0^{\rm f} \right>$ of $H_{\rm f}$
up to corrections of third order:
\begin{equation}
\overline{\left< O \right>} = 2 \left< E_0^{\rm f} \right| O \left| E_0^{\rm f} \right> + 
{\mathcal{O}}(g^3) .
\label{moeckel}
\end{equation}
To prove this theorem it was assumed that:
\begin{enumerate}
 \item[(a)] Perturbation theory in $H_{\rm f}-H_{\rm i}$ is applicable for the eigenstates. 
\item[(b)] All discrete (finite system size) eigenenergies of $H_{\rm i}$ 
and $H_{\rm f}$ are nondegenerate.
\item[(c)] A basis of common eigenstates of $H_{\rm i}$ and $O$ exists, i.e., $[O,H_{\rm i}]=0$.
\item[(d)] The initial state is the ground state $\left| E_0^{\rm i} \right>$  with respect to 
$H_{\rm i}$.
\item[(e)]  $O \left| E_0^{\rm i} \right>=0$.
\end{enumerate}
Note that under these conditions, $\overline{\left< O \right>}$ is at least of 
order $g^2$ (there is no linear term).

Out of the above conditions, (e) is not restrictive as it can always be 
achieved via a redefinition of the observable by subtracting its expectation value in the state 
$\left| E_0^{\rm i} \right>$ (see below). More importantly, (c) limits the class of operators to which the theorem applies. The standard example studied in Refs.~\onlinecite{Moeckel08} and  
\onlinecite{Moeckel09} is the nonlocal eigenmode number operator with respect to 
$H_{\rm i}$, which leads to the momentum distribution function if $H_{\rm i}$ is given 
by the kinetic energy (interaction quench out of the noninteracting ground state). 

If one now assumes that 
\begin{equation}
\left< O \right>_{\rm ss} = 
\lim_{t \to \infty} \lim_{L \to \infty} \left< E_0^{\rm i} \right| O(t)\left| E_0^{\rm i} \right>
\label{stationary}
\end{equation}
converges to a steady-state value, it is tempting to conclude that this and the 
time average $\overline{\left< O \right>}$ agree. This, however, ignores potential 
subtleties when performing the thermodynamic limit $L\to\infty$. (While our general considerations hold in any spacial dimension, all case studies focus on 1d models; hence, we use the symbol $L$ for the system size throughout 
our paper.) In a closed system, a steady state can only be reached if the thermodynamic limit $L\to\infty$ is 
taken before $t\to\infty$ to prevent recurrence effects -- this explains the 
order of limits in Eq.~(\ref{stationary}). However, in order to compare  
$\overline{\left< O \right>}$ with $\left< O \right>_{\rm ss}$, Eq.~(\ref{blarichtig}) must be evaluated for $L\to\infty$; the limits of large times and 
large systems are therefore effectively reversed. In general, it is not clear that this 
swapping is permitted. Even worse, the time average might lead to a well-defined 
result while the right-hand side of Eq.~(\ref{stationary}) does not even converge. 
Given this plethora of subtleties, we will employ our insights for time averages only to guide our intuition of what to expect for the steady state.     

Under this caveat, we still generalize the theorem of 
Refs.~\onlinecite{Moeckel08} and  \onlinecite{Moeckel09} on the time average of expectation values by loosening (b), (c), and (d). In particular, we allow for eigenstates to be degenerate and consider observables which do not necessarily commute with $H_{\rm i}$. Subsequently, we develop a phenomenology
of thermalization for small quenches. For the rest of the paper, we introduce the notation
\begin{eqnarray}
H_{\rm f}= H_{\rm i} + g V.
\label{gVdef}
\end{eqnarray}

{\it Generalizations ---} Throughout this paper, we abandon the assumption of nondegenerate 
spectra. In fact, the standard many-body models studied in the realm of quantum quenches generically feature degeneracies. This can already be seen for a 1d tight-binding chain with periodic boundary conditions, which is one prototypical choice of $H_{\rm i}$. Many-body eigenstates of this system are generically degenerate, partly due to the translation invariance.

Guided by the idea that the initial and final Hamiltonian share the same 
symmetries responsible for the degeneracies (e.g., translation invariance), 
we replace the assumption (b) by the weaker one
\begin{enumerate}
 \item [(b2)] The eigenstates $\left| E_n^{\rm i},\lambda \right>$ of 
$H_{\rm i}$  and $\left| E_n^{\rm f},\lambda \right>$ of $H_{\rm f}$ (with $n \in {\mathbb N}_0$) 
can be characterized by the same (set of) additional 
quantum number(s) $\lambda$ fully lifting possible 
degeneracies (complete set of commuting observables).
\end{enumerate}
If, e.g., $H_{\rm i}$ and  $H_{\rm f}$ are invariant under translations, the total (lattice) momentum is one component of $\lambda$. (a) and (b2) are the central assumptions which are exploited in all our model-independent considerations.

Moreover, we loosen (d) and allow all eigenstates  $\left| E_{l}^{\rm i},\lambda \right>$ 
of $H_{\rm i}$ as initial states (not only a nondegenerate $\left| E_{0}^{\rm i}\right>$). As mentioned above, assumption (e) is not restrictive; a trivial $g$-independent term can be eliminated by considering $\tilde O = O - \left< E_{l}^{\rm i},\lambda \right| O \left| E_ {l}^{\rm i},\lambda \right> $. For brevity of notation, we drop the tilde 
and assume that the initial state expectation value of $O$ was 
subtracted; we make this explicit by reintroducing the tilde whenever 
appropriate. Note that in our approach $O$ can be a self-adjoint operator (observable) or the operator 
part of a correlation function. For convenience we always refer to $O$ as an 
observable.

We then discuss several generalizations of the class of observables to which the theorem applies. In a first step (see Sect.~\ref{sec:general1st}), we drop (c) completely and -- using only (a) and (b2) -- prove that for arbitrary $O$
\begin{equation}
\overline{\left< O \right>} = \left< E_{l}^{\rm f},\lambda \right| 
O \left| E_{l}^{\rm f},\lambda \right> + {\mathcal{O}}(g^2)
\label{firstorder}
\end{equation} 
holds. Up to linear order in the quench parameter $g$, the time-averaged expectation value is 
thus equal to the expectation value in the eigenstate $\left| E_{l}^{\rm f},\lambda \right>$ 
of $H_{\rm f}$. Note, however, that the linear term is not necessarily non-zero for each observable.

In a second step (see Sect.~\ref{sec:factor2}), we go up to second order in $g$ as in 
Eq.~(\ref{moeckel}) but weaken the condition (c). Instead, we assume that
\begin{enumerate}
 \item[(c2)] $O$ is chosen such that 
\begin{equation}
\left< E_{n}^{\rm i},\lambda \right| O \left| E_{m}^{\rm i},\lambda \right> 
= \delta_{n,m} \left< E_{n}^{\rm i},\lambda \right| O \left| E_{n}^{\rm i},\lambda \right> 
\label{weaker}
\end{equation}
holds for all $n$, $m$, and $\lambda$.
\end{enumerate}
This means that for any given (set of) additional quantum number(s) $\lambda$, the observable does not 
couple energy-subspaces of $H_{\rm i}$. 
Under the conditions (a), (b2), and (c2), we can then prove that 
\begin{equation}
 \overline{\left< O \right>} = 2 \left< E_{l}^{\rm f},\lambda \right| O \left| E_{l}^{\rm f},\lambda \right> + 
{\mathcal{O}}(g^3)
\label{moeckelallg}
\end{equation}
holds. The linear order in $g$ vanishes in Eq.~(\ref{moeckelallg}). Note that the assumption (c) of Ref.~\onlinecite{Moeckel09}  implies (c2) but not the converse; our Eq.~(\ref{moeckelallg}) is the natural generalization of Eq.~(\ref{moeckel}). We thus extend the class of operators for which the time-averaged expectation  
value can be expressed in terms of an eigenstate expectation value 
of $H_{\rm f}$  up to order $g^2$. This will turn out to be crucial as we 
are mainly interested in observables which are 
spatially local and hence do not satisfy (c).

The obvious questions are: Are the assumptions (b2) and (c2) fulfilled in realistic setups? 
To this end, is it straightforward to identify the set of $\lambda$ 
for given standard $H_{\rm i/f}$, and can all degeneracies be lifted by 
taking into account quantum numbers associated to fundamental symmetries (e.g., translation invariance)? As our 
case studies of Sects.~\ref{sec:tightbinding}-\ref{sec:lattice} illustrate, 
both must be answered by `no'. This can already be seen in the 1d tight-binding
chain, which, besides the degeneracies associated with the translational symmetry, features further
`accidental' degeneracies.\cite{Yuzbashyan02}

At this stage, it might thus remain fuzzy 
how results for time-averaged expectation values based on the above assumptions can
be useful in understanding the large-time quench dynamics of such Hamiltonians. 
However, this will become obvious 
in the course of the paper: They can be employed to develop a phenomenology which 
applies to the broad range of  models and local observables of interest to us.

\subsection{Thermalization}

We will eventually test our (to be developed) `small-quench thermalization phenomenology' 
by case studies where we compute the time evolution for representative 
1d fermionic models from four different classes and a variety of 
observables (Sects.~\ref{sec:tightbinding}-\ref{sec:lattice}). The 
classes are (i) quadratic Hamiltonians, (ii) Hamiltonians 
which can be rewritten as quadratic forms in effective degrees of freedom, 
(iii) nonquadratic Hamiltonians which are characterized by an 
extensive set of (quasi-) local integrals of motion (Bethe ansatz solvable), 
and (iv) those for which no such set is known (and believed to be 
nonexistent). We explicitly verify that the expectation value of all 
observables of interest to us approach a stationary value at large times. 

For quadratic or effectively quadratic $H_{\rm f}$, the eigenmode occupancies 
with respect to $H_{\rm f}$ constitute a set of (nonlocal) integrals of motion. 
In the steady state, the occupancies are thus not distributed according to 
equilibrium Fermi-Dirac or Bose-Einstein statistics but instead determined 
by their initial-state expectation values; they do not thermalize. One might 
still wonder if for such models and large times the expectation values of certain 
observables (e.g., {\it local} ones) are equal to their thermal counterparts at an 
appropriately chosen temperature $T_{\rm f}$. If so, this is believed to be even more 
likely for nonquadratic models.
 
In fact, this idea was put forward in Ref.~\onlinecite{Berges04}.
It was argued and exemplified that {\it local} observables reach their thermal 
steady-state values by dephasing largely independent of the dynamics of the mode 
occupancies of the quasiparticles.  
Thus a thermal distribution of the nonlocal mode occupancies might not be necessary
for other 
observables to be effectively thermal. To quote from Ref.~\onlinecite{Berges04}: 
``Different quantities effectively thermalize on different time scales and a complete 
thermalization of all quantities may not be necessary.'' In nonquadratic models 
quasiparticle scattering starts to affect the dynamics of the mode occupancies
on scales which are large compared to those on which the local observables 
reach their thermal value. Depending on the model it might eventually lead 
to thermal steady-state values. Before scattering sets in, the mode occupancies 
are stuck in a prethermalization plateau.  For weakly-perturbed quadratic Hamiltonians, 
the occupancies in this time regime can be described by a generalized Gibbs 
ensemble (GGE).\cite{Kollar11} To distinguish the fast thermalization of local 
observables from the slow relaxation of the mode occupancies (in nonquadratic models) -- 
which might or might not lead to thermal expectation values of the latter -- 
the authors of Ref.~\onlinecite{Berges04} denoted the suggested scenario as {\it 
prethermalization.} In the following we refer to it as the {\it prethermalization conjecture.}
We emphasize that the prethermalization conjecture does {\it not} imply that {\it local} 
observables, after quickly reaching a time-independent value by dephasing, show 
further changes at larger time scales when quasiparticle scattering starts to affect 
the dynamics of the mode occupancies in nonquadratic models.\cite{Berges04,Eckstein09} 
Consistently, such a behavior was to the best of our knowledge not observed in 
model calculations.  Here, we do not investigate the dynamics of the mode occupancies 
(of nonquadratic models) and do thus not address the question whether these take thermal 
values or not.

\subsection{Thermalization for small quenches}

For small quenches, we can derive model-independent results on thermalization (see Sect.~\ref{sec:general}). In conjunction with the insights on time averages, this allows us to formulate our `phenomenology of small quenches'. Supplementing (a) and (b2), we now make the additional assumption that
\begin{enumerate}
 \item[(f)] The ground state $\left| E_0^{\rm i},\lambda_0 \right> = 
\left| E_0^{\rm i}\right> $ of $H_{\rm i}$ is nondegenerate, and the system is initially prepared in this state.
\end{enumerate}
Here $\lambda_0$ denotes the value(s) of the additional quantum number(s) taken in the ground state. 
The condition (f) will be fulfilled in all case studies of Sects.~\ref{sec:tightbinding}-\ref{sec:lattice} [classes (i) to (iv)].

{\it Effective temperature in gapless systems --- } In Sect.~\ref{sec:generaltemp}, we present 
results for the dependence of $T_{\rm f}$ on $g$ in {\it metallic (`gapless')} systems. 
The effective temperature $T_{\rm f}$ 
is chosen such that the canonical (or grand canonical; see below) expectation value 
of $H_{\rm f}$ equals the energy quenched into the system,
\begin{eqnarray}
\left< E_0^{\rm i} \right| H_{\rm f} \left| E_0^{\rm i} \right> 
\overset{!}{=} \frac{1}{Z_{\rm f}} \mbox{Tr} \left( e^{-\beta_{\rm f } H_{\rm f}} H_{\rm f}\right) 
= \left< H_{\rm f} \right>_{\rm th}, 
\label{Teffdef}
\end{eqnarray}
where $Z_{\rm f}$ is the partition function with respect to the final Hamiltonian, $\beta_{\rm f} = T^{-1}_{\rm f}$ denotes the inverse temperature, and  
we introduced the notation $ \left< \dots \right>_{\rm th}$ for the thermal expectation value. 
We show that if both sides of Eq.~(\ref{Teffdef}) are expanded to order $g$, one obtains $T_{\rm f} =0$; an expansion to order $g^2$ leads to $T_{\rm f} \propto g$. This {\it suggests} that the following identity holds:
\begin{enumerate}
 \item[(g)]
 \begin{equation}\label{Othgs}
 \left< O \right>_{\rm th} = \left< O \right>_{\rm th}(T_{\rm f}=0) + {\mathcal{O}}(g^2)  
 \end{equation}
\end{enumerate}
for any given $O$.

{\it Thermalization to first order --- } In Sect.~\ref{sec:firstordertherm}, we argue that if 
we assume the conditions (a), (b2), (f), and (g) to hold and moreover read Eq.~(\ref{firstorder}) 
as an equation for the steady state (and not only the time average), every observable which 
becomes stationary thermalizes to linear order in $g$ (note, however, that the linear term is 
not necessarily finite for each $O$). We emphasize  that this statement holds true even for 
(effectively) quadratic Hamiltonians and that it is independent of any specific assumptions on the 
locality of the observable. Note that by speaking of `thermalization' when a 
steady-state expectation value becomes equal to the ground state (with respect to the final 
Hamiltonian) one we might extend the meaning of this phrase, which frequently appears to be 
reserved for cases in which $T_{\rm f} >0$. However, we believe that this is meaningful as 
in case studies `first order thermalization' will turn out to be a frequently encountered 
phenomenon.

In our case studies of Sects.~\ref{sec:tightbinding}-\ref{sec:lattice}, we will explicitly 
confirm this `first order thermalization' conjecture by computing $\left< O \right>_{\rm ss/th}$ 
for a variety of models and observables. We will also calculate ground-state expectation values 
to directly verify Eq.~({\ref{firstorder}), which is a central ingredient in the derivation of 
our result. Moreover, we demonstrate that the more {\it local} the observable at hand, the longer 
the term linear in $g$ prevails when increasing $g$ (and thus the longer linear order thermalization 
dominates).  

Besides its fundamental importance, this insight has direct implications for numerical 
studies. At small $g$, the linear order term dominates, rendering it rather difficult 
to observe numerical differences between the thermal expectation values and the 
steady-state ones which might appear at higher orders. Examples for this are 
discussed in Sects.~\ref{sec:tightbinding} and \ref{sec:lattice}; in fact, our case studies 
show that linear terms govern the behavior of prototypical models and {\it local} 
observables up to surprisingly large $g$. E.g., the interaction quench in the XXZ chain at half filling  is dominated by first-order thermalization over the entire gapless 
regime (see the lower three panels of Fig.~\ref{fig:comp_nI}).  
      
{\it Thermalization to second order --- }  In Sect.~\ref{sec:secondordertherm}, we discuss thermalization 
up to second order. One motivation for this is that for several observables of interest, the linear order term 
in the steady-state expectation value vanishes. We demonstrate that if the conditions that lead to first-order thermalization are satisfied, then observables with expectation values $\left< O  \right>_{\rm ss/th} = c_1 +c_2 \left< H_{\rm i} \right>_{\rm ss/th}$ (with $g$-independent $c_i \in  \mathbb{C}$) thermalize even to second order. We refer to 
this class of operators as the `thermalization class'; by definition, $H_{\rm i}$ is one of its elements ($c_1=0$ and $c_2=1$). Our explicit results of Sects.~\ref{sec:tightbinding}-\ref{sec:lattice} are fully consistent with this. Importantly, we show that many of the `standard' {\it local} observables studied in quantum quench problems are in fact members of the `thermalization class' and hence thermalize to second order in $g$. 

As a second class of operators, we define the `factor of two class', which is the natural generalization of the class of operators studied in Refs.~\onlinecite{Moeckel08} and  
\onlinecite{Moeckel09}. Its elements 
fulfill Eq.~(\ref{moeckelallg}) with the time average replaced by the steady-state 
expectation value [note that the leading term in Eq.~(\ref{moeckelallg}) is quadratic]. In Sect.~\ref{sec:secondordertherm}, we show that for `thermalization class' 
operators which are simultaneously from the `factor of two class',  the factor of 2 in
Eq.~(\ref{moeckelallg}) has a natural explanation: The steady-state 
value consists of two {\it equal} parts, the zero temperature one (ground state with 
respect to $H_{\rm f}$) as well as the one originating from thermal excitations at $T_{\rm f}$. If the conditions (a) and (b2) are fulfilled in a given system, the kinetic energy satisfies Eq.~(\ref{weaker}) and thus also Eq.~(\ref{moeckelallg}), suggesting that the `thermalization class' is in fact a subclass of the `factor of two class'. This is again corroborated by all our explicit results.

In our case studies, we also show explicitly that when leaving the `thermalization class' but  
staying in the `factor of two class' (so that no linear terms appear), the steady-state and thermal expectation values do no longer agree up to second order in the quench amplitude.
However, being order $g^2$, the steady-state as well as the thermal expectation values of such observables 
are small at small quenches (for concrete examples, see Sects.~\ref{sec:tightbinding} and \ref{sec:lattice}). 
In purely numerical studies the difference between the two is thus
easily obscured by errors inherent to the methods used and the way the steady-state value was 
extracted. Combined with our insights on linear-order
thermalization, this leads us to conclude that it is virtually impossible to make any reliable 
statements about thermalization of {\it local} observables at small quenches solely based 
on numerics. What  `small' means depends on the model, the quench parameter, as well as 
the observable (for examples, 
see Sects.~\ref{sec:tightbinding} and \ref{sec:lattice}) and will often be a priori unknown. 
    
These insights form the `thermalization phenomenology for small quenches' mentioned 
in Sect.~\ref{smallquenchintro}. It is based on our results on time averages.

\subsection{Quenches in (effectively) quadratic models: \\ classes (i) and (ii)}

Our first model system is a tight-binding chain featuring a staggered onsite energy, which we use
as the quench parameter $g$ (see Sect.~\ref{sec:tightbinding}). It represents 
the class (i) of noninteracting models with quadratic Hamiltonians.
We focus on quarter filling of the band so that the system remains metallic; the ground state at zero staggered field is chosen as the initial state. 
The time evolution can be solved exactly, and one can derive simple 
explicit expressions for steady-state expectation values of 
$G_{j,j+r} = c_j^\dag c_{j+r}$, which is one of the standard observables 
studied for quenches in lattice models ($c_j^{(\dag)}$ denote Wannier state ladder operators on 
the lattice site $j$).

We explicitely show that for even 
$r$ the steady-state expectation value of $G_{j,j+r}$ contains terms linear in $g$ 
and that Eq.~(\ref{firstorder}) with $\overline{\left< \ldots \right>}$ replaced by  $\left< \ldots \right>_{\rm ss }$ 
holds. (In the following, we will implicitly assume that this replacement was made when referring to Eqs.~(\ref{firstorder}) 
and (\ref{moeckelallg}) in the context of steady-state expectation values.) For odd $r$, the leading term of $\left< \tilde G_{j,j+r} \right>_{\rm ss}$ is quadratic, and $\tilde G_{j,j+r}$ a `factor of two class' operator [it satisfies Eq.~(\ref{moeckelallg})]. We note that for 
general $r$,  $\tilde G_{j,j+r}$  does not fall into the class of observables 
considered in Refs.~\onlinecite{Moeckel08} and \onlinecite{Moeckel09} 
as it does not commute with $H_{\rm i}$.  Since $c_j^\dag c_{j+1}$ is directly linked to the 
kinetic energy, we investigate this quantity as a byproduct. 

We derive an analytic expression 
for $T_{\rm f}$ and compute 
$\left<G_{j,j+r}\right>_{\rm th}$ up to a given order in $g$. We explicitly demonstrate that $\left<G_{j,j+r}\right>_{\rm th}=\left<G_{j,j+r}\right>_{\rm ss}+{\mathcal O}(g^2)$ holds (first order 
thermalization). For $r=1$ (locality), $\tilde G_{j,j+1}$ is simultaneously a member of the 
`factor of two class' as well as of the `thermalization class', and we accordingly observe second order thermalization, $\left<\tilde G_{j,j+1}\right>_{\rm th}=\left<\tilde G_{j,j+1}\right>_{\rm ss}+{\mathcal O}(g^3)$. 
The same holds true if $\tilde G_{j,j+1}$ is replaced by $\tilde H_{\rm i}$ (see the definition of 
the `thermalization class'). We show that for odd $r>1$, the prefactors 
of the quadratic terms of the steady-state and thermal expectation values do indeed deviate; for 
these $r$,  $\tilde G_{j,j+r}$ is no longer an element of the `thermalization class'.

Next (see Sect.~\ref{sec:TLmodel}), we investigate the translationally-invariant 
Tomonaga-Luttinger (TL) model and study interaction quenches out of its noninteracting ground state.\cite{Giamarchi03,Schoenhammer05} 
The TL model represents the class (ii) of models which can be mapped onto 
noninteracting ones with a Hamiltonian that is quadratic in effective degrees 
of freedom; in the present case, these are the bosonic densities. Closed analytic 
expressions for the quench dynamics of observables can be 
derived.\cite{Cazalilla06,Uhrig09,Iucci09,Rentrop12,Karrasch12} 

For the TL 
model, we compute the kinetic energy, the fermionic single-particle Green function, and the 
density-density correlation function. For arbitrary spatial distances of the involved operators, 
the last two quantities do not commute with $H_{\rm i}$. The steady-state expectation values simplify 
considerably for small quenches. We show that all of the above observables fall into 
the `factor of two class' and that Eq.~(\ref{moeckelallg}) is satisfied for the steady state. We derive 
explicit expressions for $T_{\rm f}$ as well as for the thermal expectation values. To leading nontrivial 
order in the spatial distance (locality), the Green function and density-density correlation function 
are a member of the `thermalization class' and thus thermalize to second order.     
We briefly discuss the semi-infinite TL model.\cite{Fabrizio95} In this case, the steady-state 
expectation value of the density has a ${\mathcal O}(g)$ contribution;  
Eq.~(\ref{firstorder}) holds, and we find thermalization to linear order.   

The results for both models are in full accordance with our 
`small-quench thermalization phenomenology'.

\subsection{Quenches in interacting lattice models: \\  classes (iii) and (iv)}
\label{classes3_4}
  
In Sect.~\ref{sec:lattice}, we go beyond Hamiltonians which can be written as a 
quadratic form and consider quenches of the two-particle interaction in a 
tight-binding model. We employ the time-dependent density-matrix renormalization 
group\cite{White92,Schollwoeck11} (DMRG) to compute the time 
evolution of the expectation value of a variety of observables. For the problem at hand, 
this numerical approach provides highly-accurate results for times up to a hundred times 
the inverse bandwidth.

In Sect.~\ref{sec:timeevol}, we first demonstrate (for different sets of parameters) that one can access time scales on which local observables approach plateau values. While one might wonder if at larger times further dynamics
sets in, at least for the special scenario where a 
so-called dimer or a N\'eel state is evolved with a $H_{\rm f}$ featuring sufficiently strong 
nearest-neighbor interactions, it was earlier shown that the steady state can indeed be 
reached with the DMRG.\cite{Fagotti14,Pozsgay14} For these protocols, exact 
results for steady-state expectation values are available as a frame of 
reference.\cite{Wouters14,Pozsgay14} 
We cannot exclude that the constant values reached for other parameters and 
initial states are not the asymptotic ones, but in our examples we do not find any 
indications of the onset of deviations from the plateau value at larger times. 
This is consistent with the prethermalization conjecture that 
local observables can become stationary by dephasing largely independent of the 
dynamics of the mode occupancies. The latter -- which we do not study as they cannot 
be computed with the same accuracy as local observables\cite{Karrasch12} -- might not 
have reached their steady-state values for the times accessible by the DMRG. We do 
not observe any systematic differences for the Bethe ansatz solvable case with nearest-neighbor 
interaction and the one in which a next-nearest-neighbor interaction is considered 
in addition; for the latter, no Bethe ansatz 
solution is known. 

We compare the expectation values obtained at the largest accessible times 
to the thermal ones (the latter are extracted via the DMRG as well). We investigate 
observables which fall into the `factor of two class', into 
the `thermalization class', and those which do not fall in any of the two classes [in order to determine the class of an operator, we explicitly check Eqs.~(\ref{firstorder}) and (\ref{moeckelallg})] . This is done 
in Sect.~\ref{sec:nearest} for the system with nearest-neighbor interaction only, 
representing the class (iii) of models  which is characterized by an extensive set 
of local and quasi-local integrals of motion without being quadratic (see e.g. 
Ref.~\onlinecite{Ilievski15}). In Sect.~\ref{sec:next}, we switch on an additional next-nearest-neighbor interaction to obtain a representative of the most general class (iv)  of models  which are not quadratic and for which no extensive set 
of local integrals of motion is known (and expected) to exist.
The numerical results turn out to be fully consistent with our thermalization phenomenology;
general observables thermalize to linear, local `thermalization class' ones to second order, respectively. We do not observe any systematic differences between the model with an extensive 
set of (quasi-) local integrals of motion and the one for which such a set is not expected to exist. 
In analogy with our results for the (effectively) quadratic models, 
the differences between thermal and steady state expectation values turn out to be generically very small 
even for sizable $g$ and {\it local} observables which do not fall into the `thermalization class'. 

\section{General results for small quenches}
\label{sec:general}

We now derive our model-independent results. Throughout this section, we assume that the conditions (a) and (b2) introduced in Sec.~\ref{smallquenchintro} hold.

\subsection{Time-averaged expectation values to first order}
\label{sec:general1st}

As our initial state we consider an eigenstate $\left|  E_{l}^{\rm i},\lambda \right>$
of $H_{\rm i}$. By inserting partitions of unity, one obtains
\begin{align}
& \overline{\left< O \right>} = 
 \overline{ \left< E_{l}^{\rm i},\lambda  \right| e^{i H_{\rm f} t} O e^{-i H_{\rm f} t} 
\left|  E_{l}^{\rm i},\lambda \right>} \nonumber \\
 = &  \lim_{\tau \to \infty} \frac{1}{\tau} \int_0^\tau \sum_{m,m'} \sum_{\kappa,\kappa'}
e^{i (E_m^{\rm f} - E_{m'}^{\rm f}) t} \left<  E_{l}^{\rm i},\lambda  \right|  \left. E_{m}^{\rm f} , \kappa  \right> \nonumber \\
& \quad \quad \quad \times \left<  E_{m}^{\rm f},\kappa  \right| O  \left| E_{m'}^{\rm f},\kappa'  \right>   
\left<  E_{m'}^{\rm f} ,\kappa' \right|  \left.  E_{l}^{\rm i} ,\lambda \right> dt \nonumber \\
= & \sum_{m} \sum_{\kappa,\kappa'} \left<  E_{l}^{\rm i} ,\lambda \right|  \left. E_{m}^{\rm f},\kappa  \right> 
\left<  E_{m}^{\rm f},\kappa  \right| O  \left| E_{m}^{\rm f},\kappa'  \right>   \nonumber \\
& \quad \quad \quad \times \left<  E_{m}^{\rm f},\kappa'  \right|  \left.  E_{l}^{\rm i} ,\lambda \right>\nonumber \\
 = & \sum_{m}  \left<  E_{l}^{\rm i},\lambda  \right|  \left. E_{m}^{\rm f},\lambda  \right> 
\left<  E_{m}^{\rm f} ,\lambda \right| O  \left| E_{m}^{\rm f},\lambda  \right>   
\left<  E_{m}^{\rm f} ,\lambda \right|  \left.  E_{l}^{\rm i},\lambda  \right> ,
\label{eq:zwischen}
\end{align}    
with $E_n^{\rm f}$ denoting the eigenvalues of $H_{\rm f}$.  
In the last step we used that
\begin{eqnarray}     
 \left<  E_{n}^{\rm i}, \lambda  \right|  \left. E_{m}^{\rm f},\kappa  \right> = 
\delta_{\lambda,\kappa} \left<  E_{n}^{\rm i}, \lambda  \right|  \left. E_{m}^{\rm f},\lambda  \right>  ,
\end{eqnarray} 
which follows from the assumption that the eigenstates of $H_{\rm i}$  and $H_{\rm f}$ share the same (set of) additional 
quantum number(s) (i.e., the same related symmetries). For any given $\lambda$, we can use standard nondegenerate 
perturbation theory to show 
\begin{eqnarray}    
 \left<  E_{n}^{\rm i},\lambda  \right|  \left. E_{m}^{\rm f},\lambda  \right> =   
\left\{  \begin{array}{ll} 
1 + {\mathcal O}(g^2) & m = n \\
g \frac{\left<  E_{n}^{\rm i},\lambda  \right| V \left| E_{m}^{\rm i},\lambda  \right> }{E_m^{\rm i} -E_n^{\rm i} } 
+ {\mathcal O}(g^2)
& m \neq  n 
\end{array} \right. . 
\label{eq:me}
\end{eqnarray}
Inserting this in Eq.~(\ref{eq:zwischen}) leads to  Eq.~(\ref{firstorder}), which completes its proof.

\subsection{Time-averaged expectation values to second order}
\label{sec:factor2}

To prove Eq.~(\ref{moeckelallg}), we  proceed as we did in the first steps of  Eq.~(\ref{eq:zwischen}) 
but insert two more partitions of unity in terms of the eigenstates of $H_{\rm i}$. This yields 
\begin{align}
 &  \overline{\left< O \right>} = 
 \sum_{m,n,n'}  \left<  E_{l}^{\rm i},\lambda  \right|  \left. E_{m}^{\rm f} ,\lambda \right>
\left<  E_{m}^{\rm f} ,\lambda \right|  \left. E_{n}^{\rm i} ,\lambda \right>
\nonumber \\
& \quad \quad \times \left<  E_{n}^{\rm i},\lambda  \right| O  \left| E_{n'}^{\rm i} ,\lambda \right> 
 \left<  E_{n'}^{\rm i},\lambda  \right|  \left. E_{m}^{\rm f} ,\lambda \right> 
\left<  E_{m}^{\rm f} ,\lambda \right|  \left.  E_{l}^{\rm i} ,\lambda \right> .
\label{eq:zwischen2}
\end{align}   
If we now focus on the special set of observables fulfilling Eq.~(\ref{weaker}) [i.e., condition (c2)],
we obtain 
 \begin{align}
 \overline{\left< O \right>}= & 
 \sum_{m, n \neq l} \left<  E_{n}^{\rm i} ,\lambda \right| O  \left| E_{n}^{\rm i} ,\lambda \right>  
 \label{noetig} \\ & \!\! \times \!\! \left|  \left<  E_{l}^{\rm i},\lambda  \right|  \left. E_{m}^{\rm f} ,\lambda \right> \right|^2 
 \! \left|  \left<  E_{n}^{\rm i},\lambda  \right|  \left. E_{m}^{\rm f},\lambda  \right>  \right|^2 \!\! . \nonumber
\end{align} 
Since we have redefined the observable by subtracting $\left<  E_{l}^{\rm i} ,\lambda \right| O  
\left| E_{l}^{\rm i} ,\lambda \right>$ (remember that we dropped the tilde and consider the initial 
state $\left|  E_{l}^{\rm i},\lambda \right>$), the term $n=l$ is excluded in the sums. 
The double sum in Eq.~(\ref{noetig}) is split into the single sum with $m=l$ and the remaining 
double sum with $n \neq l$ and $m \neq l$. Employing Eq.~(\ref{eq:me}) in the first term the first 
absolut square of the wave-function overlap provides a factor $1 + {\mathcal O}(g^2)$ (equal indices).
The remaining factor equals  $\left< E_{l}^{\rm f} ,\lambda  \right| O  
\left|  E_{l}^{\rm f} ,\lambda \right>$ which itself is of second order in $g$. 
In the second term all addends with $n \neq m$ are of order $g^4$.
To order $g^2$ this term thus reduces to a single sum ($m=n \neq l $) in which the second 
absolut square of the wave-function overlap provides a factor $1 + {\mathcal O}(g^2)$. 
To second order in $g$ the remaining factor is equal to $\left< E_{l}^{\rm f} ,\lambda  \right| O  
\left|  E_{l}^{\rm f} ,\lambda \right>$. In total this leads to Eq.~(\ref{moeckelallg}), and its proof
is complete. We note that the absolut square of the wave-function overlaps has contributions 
$g^3$ which explains the addend ${\mathcal{O}}(g^3)$ in Eq.~(\ref{moeckelallg}).
The poof explicitly shows that for an observable obeying Eq.~(\ref{weaker}), 
$\overline{\left< O \right>}$ is (at least) of second order in $g$.

\subsection{The effective temperature}
\label{sec:generaltemp}

As already emphasized in the Introduction, when discussing possible thermalization we focus
on the case in which we start in the nondegenerate ground state 
$\left| E_0^{\rm i}\right>$  of $H_{\rm i}$ [condition (f)] and consider systems which in the 
thermodynamic limit are gapless.
Anticipating that for small $g$ the effective temperature $T_{\rm f}$ is small, we perform a
low-temperature expansion of the right-hand side of the defining equation (\ref{Teffdef}) and
for $L \to \infty$ obtain
\begin{align}
\frac{\left< E_0^{\rm i} \right| H_{\rm f} \left| E_0^{\rm i}\right>}{L} 
= \frac{\left< E_0^{\rm f} \right| H_{\rm f} \left| E_0^{\rm f} \right>}{L}
+ \frac{1}{2} \gamma_{\rm f} T_{\rm f}^2 + {\mathcal O} \left( T_{\rm f}^3 \right)  ,
\label{TexpanT}
\end{align}   
where $\gamma_{\rm f} T$ is the specific heat (per volume) with respect to $H_{\rm f}$. 
Equation~\eqref{TexpanT} relates the effective temperature to the excitation energy 
$\left< E_0^{\rm i} \right| H_{\rm f} \left| E_0^{\rm i} \right> 
- \left< E_0^{\rm f}  \right| H_{\rm f} \left| E_0^{\rm f} \right>$ (excess energy 
with respect to the ground state after the quench) and the coefficient $\gamma_{\rm f}$ from 
equilibrium thermodynamics. 
For small quenches, we can furthermore expand the first term
on the right-hand side of Eq.~(\ref{TexpanT}) in powers of $g$, which leads to 
\begin{align}
&\frac{ E_0^{\rm i}}{L} + \frac{g\left< E_0^{\rm i} \right| V \left| E_0^{\rm i}\right>}{L} 
= \frac{ E_0^{\rm i}}{L} + \frac{g\left< E_0^{\rm i} \right| V \left| E_0^{\rm i}  \right>}{L} \nonumber \\
&
- \frac{g^2}{L} \sum_{m>0} \frac{\left|\left< E_m^{\rm i} ,\lambda_0 \right| V \left| E_0^{\rm i},\lambda_0\right>   
  \right|^2}{E_m^{\rm i}- E_0^{\rm i}} + \frac{1}{2} \gamma_{\rm f} T_{\rm f}^2 + {\mathcal O} \left( T_{\rm f}^3,g^3 \right)  ,
\label{TexpanTg}
\end{align}  
were we used that $V$ does not couple subspaces of different 
$\lambda$.
If we only keep terms of order $g$, we obtain
\begin{align}
T_{\rm f}=0.
\label{TistNull} 
\end{align} 
If we include the term from second order 
perturbation theory, we get
\begin{align}
T_{\rm f} = g \left[ \frac{2}{\gamma_{\rm i} L}  
 \sum_{m>0} \frac{\left|\left< E_m^{\rm i} ,\lambda_0 \right| V \left| E_0^{\rm i} ,\lambda_0 \right>   
  \right|^2}{E_m^{\rm i}- E_0^{\rm i}} \right]^{1/2},
\label{Texpansecond}
\end{align}  
where we replaced $\gamma_{\rm f}$ by $\gamma_{\rm i}$ which is consistent to this order.

\subsection{Thermalization to first order}
\label{sec:firstordertherm}
 
We just concluded that if we aim at fulfilling Eq.~(\ref{Teffdef}) (which defines the effective temperature by equating the initial and thermal energy) to order 
$g$, we have to take $T_{\rm f}=0$. To this order, the reference ensemble is thus the 
zero-temperature one. This insight is our assumption (g) of Eq.~(\ref{Othgs}). If we combine it with Eq.~(\ref{firstorder}),
(with $ \left| E_{l}^{\rm f},\lambda \right>  \to \left| E_{0}^{\rm f} \right> $), 
which in terms of the steady-state expectation value can be written as
\begin{align}
\left< O \right>_{\rm ss} = \left< O \right>_{\rm th}(T_{\rm f}=0) + {\mathcal{O}}(g^2) ,
\label{musssein}
\end{align}  
this {\it suggests} that all observables which become stationary thermalize to linear order $g$:
\begin{align}
\left< O \right>_{\rm ss} &\stackrel{*}{ =} \left< E_{0}^{\rm f} \right| 
O \left| E_{0}^{\rm f}  \right> + {\mathcal{O}}(g^2) \nonumber \\ 
& \stackrel{**}{ =} \left< O \right>_{\rm th} + {\mathcal{O}}(g^2).
\label{eq:firstordertherm}
\end{align}
All the examples considered in Sects.~\ref{sec:tightbinding}-\ref{sec:lattice} will turn 
out to be consistent with this -- we will explicitly demonstrate that the identities $*$ and $**$ are satisfied in each case.
One should emphasize that this linear order thermalization  holds independently of any specific
assumptions on the locality of $O$. However, it is important to keep in mind that not every observable has a non-zero linear contribution to $\left< O \right>_{\rm ss/th} $.

To summarize, Eq.~(\ref{eq:firstordertherm}) holds if the conditions (a), (b2), (f), and (g) are fulfilled and if Eq.~(\ref{firstorder}) can be read as an expression for the steady state. While we presented general analytical arguments suggesting that (g) is in fact a consequence of (a) and (b2), we did not succeed in strictly proving this.

\subsection{Thermalization to second order}
\label{sec:secondordertherm}

We are now in a position to investigate thermalization up to second order when further specifying 
the observable. We start out with the initial Hamiltonian as $O$. It holds
\begin{align} 
\left< H_{\rm i} \right>_{\rm ss} = &    \left< H_{\rm f} \right>_{\rm ss} - g  \left< V \right>_{\rm ss}  
\nonumber \\
= &   \left< H_{\rm f} \right>_{\rm th} - g  \left<  V \right>_{\rm ss}  \nonumber \\
= &  \left<  H_{\rm i} \right>_{\rm th} + g  \left< V \right>_{\rm th} - g  \left< V \right>_{\rm ss} .
\label{rechnung}
\end{align}  
In the second line we used Eq.~(\ref{Teffdef}) defining the effective temperature 
$T_{\rm f}$ of the thermal ensemble and 
that the expectation value of $H_{\rm f}$ is an integral of motion (under the dynamics with $H_{\rm f}$).
Employing linear-order thermalization [i.e., Eq.~(\ref{eq:firstordertherm})] for $V$,
\begin{equation}
 \left< V \right>_{\rm th}
=  \left< V \right>_{\rm ss} + {\mathcal O}(g^2),
\label{21expl}
\end{equation}
we end up with
\begin{equation}
\left< H_{\rm i} \right>_{\rm ss} =  \left< H_{\rm i} \right>_{\rm th} + {\mathcal O}(g^3) .
\label{eq:thermalization}
\end{equation}
We thus argue that the steady-state expectation value of $ H_{\rm i}$ (if it exists) agrees with 
the thermal one up to second order;  $H_{\rm i}$ thermalizes to second order. Under the condition that an observable $O$ fulfills
\begin{equation}
\left< O  \right>_{\rm ss/th} = c_1 +c_2 \left< H_{\rm i} \right>_{\rm ss/th},
\label{excondition}
\end{equation}
with $g$-independent $c_i \in  \mathbb{C}$, it is straightforward to generalize 
the above equation (\ref{rechnung}) and show that  
\begin{equation}
\left< O \right>_{\rm ss} =  \left< O \right>_{\rm th} + {\mathcal O}(g^3) .
\label{eq:thermalizationO}
\end{equation}
Equation (\ref{excondition}) defines the `thermalization class' of observables; for all of its members,  $\left< O \right>_{\rm ss}$ and $\left< O \right>_{\rm th}$ agree up to second order. Note that no additional assumptions beyond the ones used in the previous Section were made to derive Eq.~(\ref{eq:thermalizationO}) for thermalization-class observables.

At first glance, the restriction introduced in Eq.~(\ref{excondition}) appears to be rather peculiar. 
However, as we show in our case studies of Sects.~\ref{sec:tightbinding}-\ref{sec:lattice}, it is satisfied for surprisingly many of the  {\it local} observables routinely computed when studying quantum quenches. We will also explictly calculate $\left< V \right>_{\rm ss/th}$ to demonstrate that Eq.~(\ref{21expl}) generically holds. Accordingly, the explicit results for  $\left< H_{\rm i} \right>_{\rm ss/th}$ as well as for various other `thermalization class' observables will be consistent with Eq.~(\ref{eq:thermalization}).

For all systems where (a) and (b2) hold, the initial Hamiltonian $H_{\rm i}$  fulfills Eq.~(\ref{weaker}) and is therefore a member of 
the `factor of two class'. Our case studies suggest that this is generically the case and that hence the `thermalization class' is a subclass of the `factor of two class'. According to Eq.~(\ref{moeckelallg}), 
\begin{eqnarray}
\left< \tilde H_{\rm i} \right>_{\rm ss} = 2 \left< E_0^{\rm f} \right| \tilde H_{\rm i} 
\left| E_0^{\rm f} \right> + {\mathcal O}(g^3) 
\end{eqnarray}
holds. With Eq.~(\ref{eq:thermalization}) we can conclude
\begin{align}
\left< \tilde H_{\rm i} \right>_{\rm ss} & = \left< \tilde H_{\rm i} \right>_{\rm th} + {\mathcal O}(g^3) \nonumber \\
& = \left< E_0^{\rm f}  \right| \tilde H_{\rm i} 
\left| E_0^{\rm f} \right> + \mbox{($T_{\rm f}>0$)-part}   + {\mathcal O}(g^3)
\end{align}
and thus 
\begin{align}
\left< E_0^{\rm f} \right| \tilde H_{\rm i} 
\left| E_0^{\rm f} \right> = \mbox{($T_{\rm f}>0$)-part}  + {\mathcal O}(g^3) .
\end{align}
This shows that to order $g^2$ the steady state expectation value consists of two {\it equal}
parts, one given by the ground state (of $H_{\rm f}$) expectation value, the other one 
originating from the $T_{\rm f}>0$ thermal excitations. In total, this provides a natural 
explanation for the factor of 2 of Eq.~(\ref{moeckelallg}) if $O$ is given by 
$\tilde H_{\rm i}$ or, more general, an operator which is simultaneously from the 
`thermalization class' and the `factor of two class'. 
In Sects.~\ref{sec:tightbinding}-\ref{sec:lattice} we illustrate this by giving a variety of explicit examples.

\section{Quench in the noninteracting tight-binding model}
\label{sec:tightbinding}

\subsection{The Hamiltonian and its eigenstates}

As our first model to illustrate the above general considerations we study the tight-binding 
chain of $M$ lattice sites with nearest-neighbor hopping of amplitude $J=1$, (dimensionless) 
staggered onsite 
energy $\delta \geq 0$, and lattice constant $a=1$ (and thus $L=M$). 
In second quantization it is given by the quadratic Hamiltonian
\begin{align}
H = -  \sum_{j=1}^{M} \left( c_{j+1}^\dag c_j + \mbox{H.c.} \right) + \delta \sum_{j=1}^M 
(-1)^j c_j^\dag c_j . 
\label{HTB}
\end{align}   
The operator $c_j^\dag$ creates a particle in the Wannier state $\left| j \right> $.  
We assume periodic boundary conditions and thus identify the lattice sites $M+1$ and 1. 
The Wannier states  $\left\{ \left| j \right> \right\}$  form a single-particle basis.
  
For $\delta=0$ the single-particle eigenstates are given by the standard plane waves 
\begin{align}
\left| k_l \right> = \frac{1}{\sqrt{M}} \sum_{j=1}^M e^{i k_l j} \left| j \right> , - \pi \leq 
k_l=\frac{2\pi}{M} l < \pi, l \in {\mathbb Z} 
\label{eps0basis}
\end{align} 
and the eigenvalues read $\epsilon(k) = -2 \cos k$. 

The single-particle problem 
with staggered field can be solved straightforwardly as well. The eigenstates are 
\begin{align}
\left| k_l , \xi \right>_\delta = A_{\xi}(k_l) \left[ \sum_{j \, \mbox{\scriptsize odd}}   e^{ik_l j}  \left| j \right>
- \frac{d_\xi(k_l)}{\epsilon(k_l)}  \sum_{j \, \mbox{\scriptsize even}}   e^{ik_l j}  \left| j \right> 
\right],
\label{epsbasis}
\end{align}  
with $- \pi/2 \leq 
k_l=\frac{2\pi}{M} l < \pi/2, l \in {\mathbb Z}$  
from the reduced first Brillouin zone,
\begin{align}
d_\xi(k) = -\xi \sqrt{\epsilon^2(k) + \delta^2} - \delta ,
\label{dDef}
\end{align}
and the normalization constant
\begin{align}
 A_{\xi}(k) = \sqrt{\frac{2}{M}} \left\{ 1 +  \left[ \frac{d_\xi(k)}{\epsilon(k)} \right]^2 \right\}^{-1/2} .
\label{normconst}
\end{align} 
The dispersion is given by $\epsilon_{\delta,\xi} (k) = \xi \sqrt{\epsilon^2(k) + \delta^2}$.  
A gap of size $2 \delta$ opens at the boundaries of the reduced first Brillouin zone. 

The many-body 
eigenstates follow from filling the single-particle ones of ascending energy up to the 
required filling factor $\nu$. For $\delta>0$ and half filling the system is a band 
insulator. 
We focus on quarter filling $\nu=1/4$ for which the 
system remains metallic. The number of lattice sites is chosen as an odd multiple of $4$ 
to prevent a degenerate ground state; generic excited many-body states are, 
however, degenerate. It is easy to see that these degeneracies cannot be fully lifted by 
adding the lattice momentum as an additional quantum number (for $\delta>0$ the lattice momentum
with respect to the reduced Brillouin zone must be taken; the corresponding momentum operator can 
be constructed along the lines discussed in Ref.~\onlinecite{Essler05}). The remaining `accidental' 
degeneracies are at least partly associated to the $x$-axis symmetry of the dispersion 
$\epsilon_{\delta,\xi} (k)$. We were not able to identify further fundamental symmetries 
and associated quantum numbers which would allow to lift these degeneracies. Thus Eq.~(\ref{weaker}) 
cannot be exploited directly.    

As the initial state we consider the  ground state $\left| E_{0}^{\rm i} \right>$ 
with $\delta_{\rm i}=0$ while the time evolution is performed with the Hamiltonian Eq.~(\ref{HTB}) 
with $\delta_{\rm f} = \delta >0$. 
 
\subsection{The observable}

The `observable' we study is the `Green function' (or bond operator)
\begin{equation}
O = G_{j,j+r} = c_j^\dag c_{j+r} .
\label{obdef}
\end{equation}
It is routinely considered when investigating quantum quenches.
Obviously, 
\begin{align}
H_{\rm i} = - \sum_{j=1}^M \left(G_{j,j+1} + \mbox{H.c.}  \right)
\label{HGrel}
\end{align}
and we simultaneously obtain results for the initial Hamiltonian, which equals the kinetic
energy. For general $r$ the $G_{j,j+r}$ do not commute with $H_{\rm i}$ and fall out of the 
domain of observables considered in Refs.~\onlinecite{Moeckel08} and  \onlinecite{Moeckel09}. 

Computing the steady-state and thermal (including $T_{\rm f}=0$) expectation 
values we show that [see Eq.~(\ref{firstorder}) with $\overline{\left< \ldots \right>}$ replaced by 
$\left< \ldots \right>_{\rm ss}$]
\begin{align}
\left< G_{j,j+r} \right>_{\rm ss} 
= \left< E_0^{\rm f} \right| G_{j,j+r} \left| E_0^{\rm f} \right> + {\mathcal O}(\delta^2), 
\quad r \, \mbox{even}
\label{tobeconfirmedexpl1}
\end{align} 
and that this expectation value agrees to order $\delta$ with the thermal one (linear
order thermalization). 

We show explicitely that  [see Eq.~(\ref{moeckelallg}) 
with $\overline{\left< \ldots \right>}$ replaced by $\left< \ldots \right>_{\rm ss}$]
\begin{align}
\left< \tilde G_{j,j+r} \right>_{\rm ss} = 2 \left< E_0^{\rm f} \right| 
\tilde G_{j,j+r} \left| E_0^{\rm f} \right> + {\mathcal O}(\delta^3),  
\quad r \, \mbox{odd} 
\label{tobeconfirmedexpl2}
\end{align} 
implying that $\tilde G_{j,j+r}$ is  a `factor of two class' operator.

Equation~(\ref{HGrel}) and the symmetry related independence of 
$\left< G_{j,j+1} \right>_{\rm ss/th}$ on $j$ implies that $G_{j,j+1}$ in addition is 
a `thermalization class' operator. We verify that Eq.~(\ref{21expl}) holds 
for $V=\sum_{j=1}^M (-1)^j c_j^\dag c_j =\sum_{j=1}^M (-1)^j G_{j,j}$. 
Combining this with the `thermalization class' properties of 
$G_{j,j+1}$ and $H_{\rm i}$ (see Sect.~\ref{sec:secondordertherm}) 
implies thermalization of $G_{j,j+1}$ and $H_{\rm i}$ up to second order. 
We explicitely verify this comparing $\left< G_{j,j+1} \right>_{\rm ss}$
and $\left< G_{j,j+1} \right>_{\rm th}$ as well as $\left< H_{\rm i} \right>_{\rm ss}$
and $\left<  H_{\rm i} \right>_{\rm th}$.
We show that for odd $r>1$ the thermal and steady-state expectation values do not agree 
to order $\delta^2$ (`factor of two class' but not `thermalization class').

The classification of the considered operators is summarized in 
Table \ref{tabletb}.

\begingroup
\squeezetable
\begin{table}
\begin{ruledtabular}
\begin{tabular}{cccc}
  observable  &  first order therm. & fact 2 cl. & therm. cl. \\ \hline
 $G_{j,j+r}$,  $r$ even &  yes & no & no \\
  $\tilde G_{j,j+r}$, $r$ odd & first order vanishes&  yes & for $r=1$  \\
 $\tilde H_{\rm i}$ &first order vanishes  & yes &  yes 
\end{tabular}
\end{ruledtabular}
\caption{Classification of the considered operators. \label{tabletb}} 
\end{table}
\endgroup

\subsection{The time evolution and steady state}

With the eigenbasis of $H$ Eq.~(\ref{HTB}) known, it is straightforward 
to compute the initial-state expectation value of $G_{j,j+r}(t)$  for all $t>0$ employing 
\begin{align}
c_{k,\xi}(t) = e^{- i \epsilon_{\delta,\xi}(k) t} c_{k,\xi}, \quad   
c^\dag_{k,\xi}(t) = e^{i \epsilon_{\delta,\xi}(k) t} c^\dag_{k,\xi} .
\label{timeevol}
\end{align} 
At fixed $t$ the thermodynamic limit $M \to \infty$ of  $\left< E_0^{\rm i} \right| G_{j,j+r}(t) 
\left|  E_0^{\rm i} \right>$ can be taken and afterwards $t$ can be sent to infinity; the expectation value 
becomes stationary at large $t$. The asymptotic value is
\begin{align}
& \left<\tilde G_{j,j+r}\right>_{\rm ss}  \nonumber \\  
& =  \left\{  \begin{array}{ll}
 - \frac{ \delta^2}{2\pi}  \int_{-\frac{\pi}{4}}^{\frac{\pi}{4}} dk 
\frac{\cos(kr)}{4 \cos^2(k)+ \delta^2} & r \, \mbox{odd} \\
-  \frac{(-1)^j  \delta }{2 \pi} 
\int_{-\frac{\pi}{4}}^{\frac{\pi}{4}} dk 
\frac{2 \cos(kr) \cos(k)}{4 \cos^2(k)+ \delta^2} & r \, \mbox{even} 
\end{array} \right. ,
\label{ssquart}
\end{align}
with
\begin{align}
\left< E_{0}^{\rm i} \right|  G_{j,j+r} \left| E_{0}^{\rm i} \right>  & = 
\frac{1}{2 \pi} \int_{-\frac{\pi}{4}}^{\frac{\pi}{4}} dk \cos(kr) \nonumber \\ & = \left\{  \begin{array}{ll}
\frac{1}{4} 
& r=0 \\
\frac{\sin(\pi r /4)}{\pi r}
& r \neq 0 
\end{array} \right. .
\label{ssquartint}
\end{align}
The leading order contributions are obtained by taking $\delta \to 0$ in the denominator.  
We find that $\left<\tilde G_{j,j+r}\right>_{\rm ss}$ 
is order $\delta^2$ for odd $r$ and order $\delta$ for even ones.  
We note that for odd $r$ only even powers in $\delta$ contribute and for even $r$ only 
odd ones.  
For arbitrary $\delta$ the integrals in Eq.~(\ref{ssquart}) can easily be performed 
numerically.

We note in passing that $\left<\tilde G_{j,j+r}\right>_{\rm ss}$ can also be obtained 
employing a GGE.\cite{Rigol07,Barthel08,Sotiriadis14} However, we preferred to 
compute the full time evolution to prove that $\left< E_0^{\rm i} \right| \tilde G_{j,j+r}(t) 
\left|  E_0^{\rm i} \right>$ indeed becomes stationary. 

To explicitly confirm Eqs.~(\ref{tobeconfirmedexpl1}) as well as ~(\ref{tobeconfirmedexpl2}) 
and verify our expectations on 
the thermalization properties we next study the thermal expectation value
of $G_{j,j+r}$ with respect to $H_{\rm f}$ including the case of vanishing temperature.  

\subsection{The thermal expectation value}

We first have to determine the effective temperature $T_{\rm f}$ corresponding to the energy 
quenched into the system as well as  the chemical potential $\mu_{\rm f}$ ensuring the required 
filling. For $T_{\rm f}$ we here go beyond the perturbative result of Sect.~\ref{sec:generaltemp}
and for the concrete Hamiltonian Eq.~(\ref{HTB}) consider the effective temperature to all orders. 
Both $T_{\rm f}$ and $\mu_{\rm f}$ are obtained by solving the set of equations
\begin{align} 
& \lim_{M \to \infty} \frac{\left< E_0^{\rm i} \right| H_{\rm f} \left|   E_0^{\rm i} \right>}{M} 
\overset{!}{=} \!\! \lim_{M \to \infty} 
\frac{1}{M Z_{\rm f}} \mbox{Tr} \left( e^{-[H_{\rm f}- \mu_{\rm f} N]/T_{\rm f}} H_{\rm f} \right) \nonumber \\  
\Leftrightarrow & - \frac{\sqrt{2}}{\pi}
\overset{!}{=}\frac{1}{2 \pi} 
\int_{-\frac{\pi}{2}}^{\frac{\pi}{2}} dk \sum_{\xi = \pm}   \epsilon_{\delta,\xi}(k) 
f \left( \frac{\bar \epsilon_{\delta,\xi}(k)}{T_{\rm f}} \right) ,
\label{TdefTB}
\end{align}
where we used that $H_{\rm f}$ can be expressed in terms of $G_{j,j}$ and $G_{j,j+1}$ as well as 
Eq.~(\ref{ssquartint}) and
\begin{align} 
& \lim_{M \to \infty} \frac{\left< E_0^{\rm i} \right| N \left|   E_0^{\rm i} \right>}{M} 
\overset{!}{=} \!\! \lim_{M \to \infty} 
\frac{1}{M Z_{\rm f}} \mbox{Tr} \left( e^{-[H_{\rm f}- \mu_{\rm f} N]/T_{\rm f}} N \right) \nonumber \\  
\Leftrightarrow & 
\nu \overset{!}{=} \frac{1}{2 \pi}  
\int_{-\frac{\pi}{2}}^{\frac{\pi}{2}} dk   \sum_{\xi = \pm} f \left( \frac{\bar \epsilon_{\delta,\xi}(k)}{T_{\rm f}} \right),
\label{mudefTB}
\end{align}
with $\bar \epsilon = \epsilon - \mu_{\rm f}$ and the Fermi function  
$f(x)= \left[  e^{x} +1   \right]^{-1}$.  
The coupled equations (\ref{TdefTB}) and (\ref{mudefTB}) can straightforwardly be solved 
numerically. For small quenches the solutions are consistent with Eqs.~(\ref{TistNull}) and 
(\ref{Texpansecond}).

For a given $T_{\rm f}$ and $\mu_{\rm f}$ the expectation value 
$\left<\tilde G_{j,j+r}\right>_{\rm th}$ in the grand canonical ensemble can 
in analogy to the time evolution be computed by appropriate basis changes.  
It is given by
\begin{align}
\left<\tilde G_{j,j+r}\right>_{\rm th} \!\!\! = &  \frac{1}{\pi} \!\! \int_{-\frac{\pi}{2}}^{\frac{\pi}{2}} \!\! dk 
\frac{\cos(kr) \cos(k)}{\sqrt{4 \cos^2(k)+ \delta^2}} 
\!\!  \sum_{\xi = \pm} \! (-\xi) f \left( \frac{\bar \epsilon_{\delta,\xi}(k)}{T_{\rm f}} \right) \nonumber \\
& - \frac{1}{2 \pi} \int_{-\frac{\pi}{4}}^{\frac{\pi}{4}} dk
\cos(kr)
\label{thodd}  
\end{align}
for $r$ odd and 
\begin{align}
& \left<\tilde G_{j,j+r}\right>_{\rm th} =
\frac{1}{2 \pi}  \int_{-\frac{\pi}{2}}^{\frac{\pi}{2}} dk \cos(kr)
\sum_{\xi = \pm} f \left( \frac{\bar \epsilon_{\delta,\xi}(k)}{T_{\rm f}} \right) \nonumber \\
& -\frac{(-1)^j}{2 \pi} \delta \int_{-\frac{\pi}{2}}^{\frac{\pi}{2}} dk \frac{\cos(kr) }{\sqrt{4 \cos^2(k)+ \delta^2}}
\sum_{\xi = \pm} (-\xi) f \left( \frac{\bar \epsilon_{\delta,\xi}(k)}{T_{\rm f}} \right) \nonumber \\
& - \frac{1}{2 \pi} \int_{-\frac{\pi}{4}}^{\frac{\pi}{4}} dk \cos(kr) 
\label{theven}
\end{align}
for $r$ even. The integrals can easily be performed numerically.   

\subsection{Comparison}
\label{sec:sec}

We are now in a position to {\it explicitly}
confirm Eqs.~(\ref{tobeconfirmedexpl1}) and (\ref{tobeconfirmedexpl2}), 
that is Eqs.~(\ref{firstorder}) and (\ref{moeckelallg}) with the time average replaced 
by the steady state expectation value. 
To achieve this we set $T_{\rm f}=0$ in Eqs.~(\ref{thodd}) and (\ref{theven}) 
and thus consider 
\begin{eqnarray}
\left<\tilde G_{j,j+r}\right>_{\rm th} (T_{\rm f}=0) = 
\left< E_0^{\rm f} \right| \tilde G_{j,j+r} \left| E_0^{\rm f} \right> . 
\end{eqnarray}
Expanding in $\delta$ gives 
\begin{align}
 & \left< E_0^{\rm f} \right|  \tilde G_{j,j+r} \left| E_0^{\rm f} \right> \nonumber \\
 & = \left\{  \begin{array}{ll}
\!\!\! - \frac{\delta^2}{2\pi}  \int_{-\frac{\pi}{4}}^{\frac{\pi}{4}} dk 
\frac{\cos(kr)}{8 \cos^2(k)} + {\mathcal O}(\delta^4) & r \, \mbox{odd} \\
\!\!\! -  \frac{(-1)^j \delta }{2 \pi}  
\int_{-\frac{\pi}{4}}^{\frac{\pi}{4}} dk 
\frac{\cos(kr)}{2 \cos(k)} + {\mathcal O}(\delta^3)  & r \, \mbox{even} 
\end{array} \right. .
\label{ssquart1}
\end{align}
Up to a factor $1/2$ for odd $r$ this agrees with the leading order 
expansion of Eq.~(\ref{ssquart}) which confirms  Eqs.~(\ref{tobeconfirmedexpl1}) 
and (\ref{tobeconfirmedexpl2}).

To show thermalization to order $\delta$ we expand  Eqs.~(\ref{thodd}) and
(\ref{theven}) to this order, taking into account the expansion of $T_{\rm f}$ 
Eq.~(\ref{Texpansecond}).
For odd $r$,  the linear contribution of $\left<\tilde G_{j,j+r}\right>_{\rm th}$
vanishes. The same holds for the steady state value $\left<\tilde G_{j,j+r}\right>_{\rm ss}$ 
of Eq.~(\ref{ssquart}).
For even $r$ we find
\begin{align}
\left<\tilde G_{j,j+r}\right>_{\rm th} =  -  \frac{(-1)^j \delta }{2 \pi}  
\int_{-\frac{\pi}{4}}^{\frac{\pi}{4}} dk \frac{\cos(kr)}{2 \cos(k)} + {\mathcal O}(\delta^2)
\label{linordertherm}
\end{align}  
which agrees with the linear order expansion of the lower line of Eq.~(\ref{ssquart}). 
We thus confirmed first order thermalization of  $G_{j,j+r}$ for arbitrary $r$ (no restriction
on the locality). 

Going to higher orders in the expansion it is straightforward to 
show that for general even $r$, $\left<\tilde G_{j,j+r}\right>_{\rm th}$ has a correction 
of order $\delta^2$. As noted in connection with Eq.~(\ref{ssquart}) the $\delta^2$ contribution 
vanishes  in $\left<\tilde G_{j,j+r}\right>_{\rm ss}$ and the agreement between the steady-state
and thermal expectation values is thus restricted to the leading order. The 
higher-order expansion of $\left<\tilde G_{j,j+r}\right>_{\rm th}$ also shows that if $r$ 
is a multiple of $4$ the $\delta^2$ term vanishes in accordance with 
$\left<\tilde G_{j,j+r}\right>_{\rm ss}$. This higher order agreement is, however, 
particular to the symmetries of the model as well as the observable considered and cannot be expected
in other cases. When comparing numerical results for  $\left<\tilde G_{j,j+r}\right>_{\rm ss}$ and  
$\left<\tilde G_{j,j+r}\right>_{\rm th}$ for even $r$ we will further down 
restrict ourselves to odd multiples of $2$ which show the generic behavior.   
 
Equation (\ref{linordertherm}) also implies that Eq.~(\ref{21expl}) holds for 
$V=\sum_{j=1}^M (-1)^j G_{j,j}$; thermal excitations for $T_{\rm f}>0$ only contribute to order 
$\delta^2$. According
to the considerations of Sect.~\ref{sec:secondordertherm} the
kinetic energy $H_{\rm i}$  as well as $G_{j,j+1}$ (`thermalization class' operator) 
thus thermalize up to order $\delta^2$. This can be seen {\it explicitly} by 
expanding Eq.~(\ref{thodd}) for $r=1$  in $\delta$ 
employing the second order expression for $T_{\rm f}$.
This gives 
\begin{align}
\left<\tilde G_{j,j+1}\right>_{\rm th}   = & 
\underbrace{- \frac{\delta^2}{2\pi}  \int_{-\frac{\pi}{4}}^{\frac{\pi}{4}} dk 
\frac{1}{8 \cos(k)}}_{T_{\rm f}=0-\mbox{\scriptsize contribution}} \nonumber \\
& \underbrace{- \frac{\delta^2}{2\pi}  
\int_{-\frac{\pi}{4}}^{\frac{\pi}{4}} dk \frac{1}{8 \cos(k)}}_{T_{\rm f}>0-\mbox{\scriptsize contribution}} 
+  {\mathcal O}(\delta^3) . 
\end{align}
The first term is the zero temperature contribution -- compare to Eq.~(\ref{ssquart1}) 
for $r=1$ -- while the 
second one stems from thermal excitations at $T_{\rm f}$; they add up to give the steady state result 
Eq.~(\ref{ssquart}) (upper right-hand side for $r=1$ and $\delta \to 0$ in the denominator).   
In accordance with our considerations 
of Sect.~\ref{sec:secondordertherm} both addends are equal. This explains why,
to order $\delta^2$,  
$\left<\tilde G_{j,j+1}\right>_{\rm ss} = \left<\tilde G_{j,j+1}\right>_{\rm th}$ 
is twice as large as  $\left< E_0^{\rm f} \right| \tilde G_{j,j+1} \left| E_0^{\rm f} \right>$.  

\begin{figure}[t]
\centering
\includegraphics[width=\columnwidth]{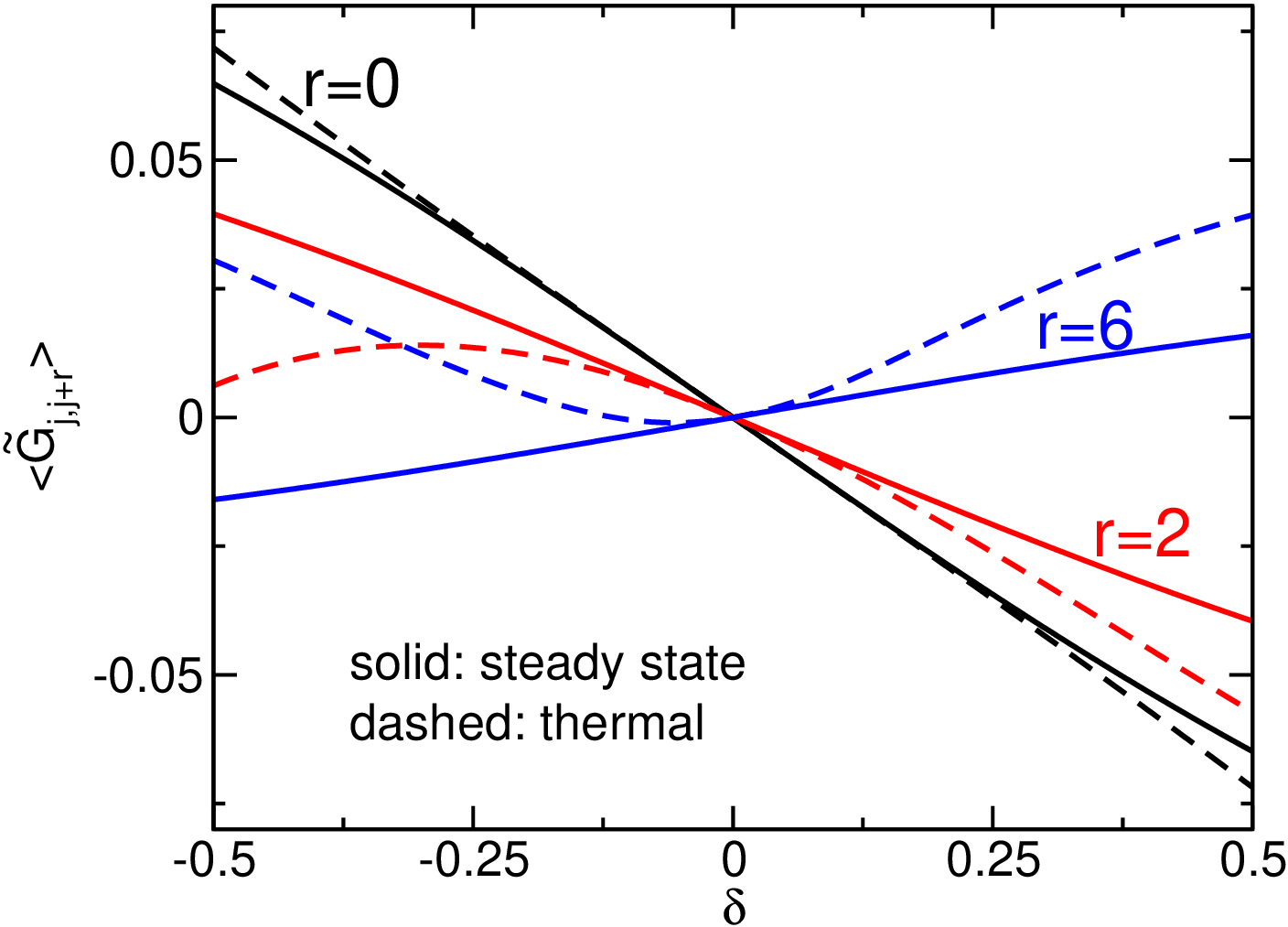}
\caption{Comparison of the steady-state and thermal expectation values 
of $\tilde G_{j,j+r}$ as a function of the quench amplitude $\delta$ for $j=0$ and different even $r$.
At fixed $r$ the two values agree to order $\delta$ (leading order thermalization). 
The more local the observable the longer the linear term prevails, that is the longer the 
steady-state and thermal values agree, when $|\delta|$ is increased.  
}
\label{fig:even_pap}
\vspace{1.0cm}
\includegraphics[width=\columnwidth]{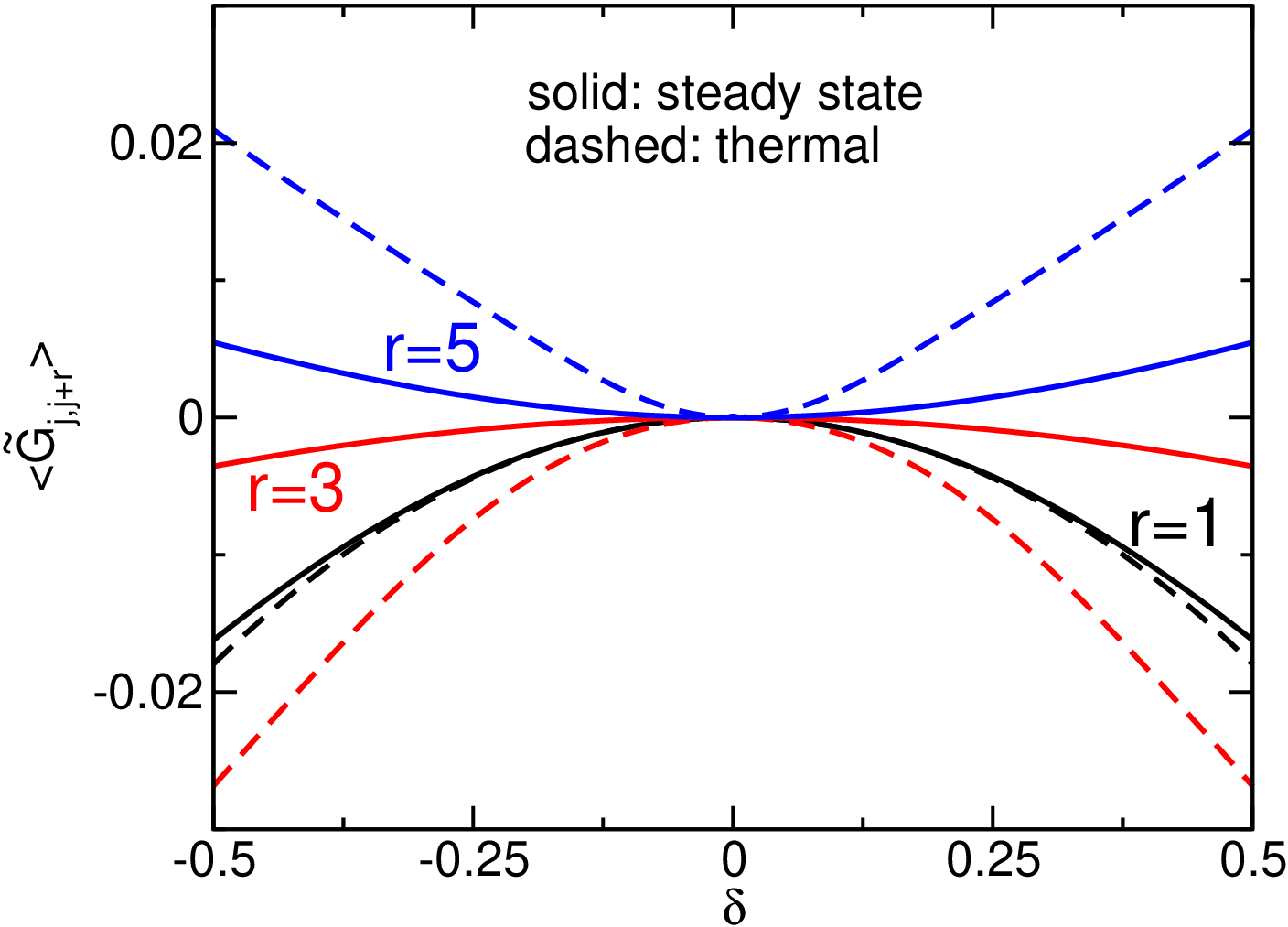}
\caption{The same as in Fig.~\ref{fig:even_pap} but for odd $r$. 
The leading $\delta$-dependence is of second order (`factor of two class' 
observables). For $r=1$ the prefactors of the second order terms agree and we find 
second order thermalization (`thermalization class' operator). This does not hold 
for $r=3,5$ as further analyzed in Fig.~\ref{fig:Delta_pap}. The more  
 local the observable the longer the 
steady-state and thermal values agree when $|\delta|$ is increased.
}
\label{fig:odd_pap}
\end{figure}

\begin{figure}[t]
\centering
\includegraphics[width=.93\columnwidth]{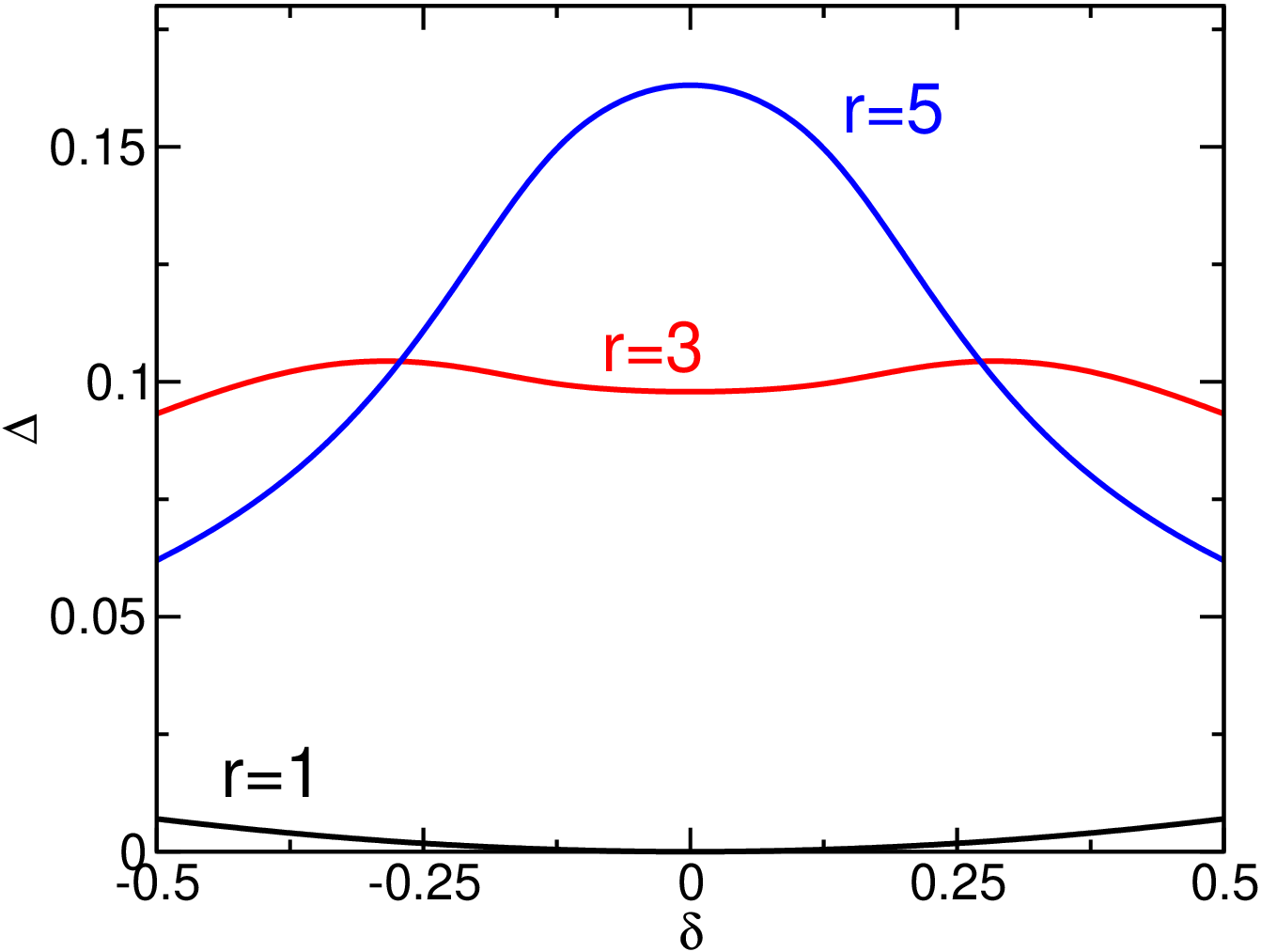}
\caption{ The difference $\Delta = 
\left|\left<\tilde G_{j,j+r}\right>_{\rm ss} - \left<\tilde G_{j,j+r}\right>_{\rm th} 
\right|/\delta^2$ as a function of $\delta$ for $j=0$ and different odd $r$.
The vanishing of $\Delta$ for $\delta \to 0$ and $r=1$ shows that 
$\tilde G_{j,j+1}$ is a `thermalization class' operator. For odd $r>1$, 
$\tilde G_{j,j+r}$ is not from this class but still from the 
`factor of two class' (see Fig.~\ref{fig:odd_pap}).}
\label{fig:Delta_pap}
\end{figure}

We finally compare results for $\left<\tilde G_{j,j+r}\right>_{\rm ss}$ and 
$\left<\tilde G_{j,j+r}\right>_{\rm th}$ obtained by numerically solving the integrals 
in the exact expressions Eqs.~(\ref{ssquart}) to (\ref{theven}). 
Figure \ref{fig:even_pap} shows $\left<\tilde G_{j,j+r}\right>_{\rm ss/th}$ 
as a function of $\delta$ for $r=0,2,6$. Consistent with our analytical insights the 
steady-state and thermal expectation values agree to linear order (first order 
thermalization). We observe that the smaller $r$, that is the more local the observable, 
the longer the linear term prevails for increasing $|\delta|$.   
Figure \ref{fig:odd_pap} shows $\left<\tilde G_{j,j+r}\right>_{\rm ss/th}$ as a function of $\delta$ 
for $r=1,3,5$. The just described trend on the locality of the observable and the 
agreement of the steady-state and thermal expectation values obviously also holds 
for odd $r$. For such $r$, $\tilde G_{j,j+r}$ is a `factor of two class' operator 
and the steady-state expectation value is ${\mathcal O}(\delta^2)$. As just shown 
analytically, the same holds for $\left<\tilde G_{j,j+r}\right>_{\rm th}$. 
This is consistent with our numerical results. 
Being order $g^2$ the expectation values for small $|\delta|$ are rather small.   
To further investigate 
the second order term in Fig.~\ref{fig:Delta_pap} we show $\Delta = 
\left|\left<\tilde G_{j,j+r}\right>_{\rm ss} - \left<\tilde G_{j,j+r}\right>_{\rm th} 
\right|/\delta^2$ for the data of Fig.~\ref{fig:odd_pap}. 
For $r=1$, $\tilde G_{j,j+r}$ is a `thermalization class' operator and 
$\lim_{\delta \to 0} \Delta =0$. In contrast, for  $r=3,5$, $\tilde G_{j,j+r}$  
is not from this class and $\lim_{\delta \to 0} \Delta$ remains finite; for odd $r>1$, 
$\tilde G_{j,j+r}$ does thus not thermalize to second order. 

Our computations exemplify that for small even or odd $r$, that is a local $\tilde G_{j,j+r}$, 
and small $|\delta|$, either the differences or the absolute values of the expectation 
values are rather small. It indicates 
that based on purely numerical data obtained for models which cannot be solved 
exactly and which are prone to errors -- for examples see Sect.~\ref{sec:lattice} -- it 
will be very difficult to make any definite statements on thermalization for small 
quenches.

\section{Interaction quench in the Tomonaga-Luttinger model}
\label{sec:TLmodel}

In this section we study steady-state expectation values of a variety of 
observables after small interaction quenches in the spinless TL model.\cite{Giamarchi03,Schoenhammer05} 
The quench dynamics of the TL model was studied before,\cite{Cazalilla06,Uhrig09,Iucci09,Rentrop12,Karrasch12} 
however, not in the 
context of interest to us. This continuum model presents one of the rare examples in which 
closed analytical expressions for the expectation values of observables and correlation 
functions at all times can be obtained for an interacting system. To achieve this 
bosonization is used. This method consists of two steps. First the fermionic 
Hamiltonian is rewritten in terms of collective bosonic degrees of freedom (bosonization 
of the Hamiltonian). This is possible as right- and 
left-moving fermions with linear dispersion as well as two-particle scattering processes with only 
small momentum transfer are considered. To be able to compute all fermionic expectation values of interest, in 
a second step one has to express the fermionic field in terms of the bosons 
(bosonization of the field operator).\cite{Schoenhammer05,vonDelft98}  
We here adopt the notation and conventions of Ref.~\onlinecite{Rentrop12}. 

\subsection{The Hamiltonian and eigenstates}

The bosonized Hamiltonian is given by
\begin{align} 
H  =  \sum_{n >0 } & \left[  k_n \left( v_{\rm F} + \frac{v(k_n)}{2 \pi} \right) 
\left( b_n^\dag b_n^{} + b_{-n}^\dag b_{-n}^{} \right) \right. \nonumber \\
& \left. + k_n   \frac{v(k_n)}{2 \pi} 
\left( b_n^\dag b_{-n}^\dag + b_{-n}^{} b_{n}^{} \right)  \right] ,
\label{TLmodel}
\end{align}
with bosonic operators $b_n^{(\dag)}$, $n \in {\mathbb Z}$,  which are linked to the 
densities $\rho_{n,\pm}$ of the right($+$)- and left($-$)-moving 
fermions in momentum space by
\begin{eqnarray}
b_n = \frac{1}{\sqrt{|n|}} \left\{ \begin{array}{cc} 
\rho_{n,+} & \mbox{for} \; n >0 \\
\rho_{n,-} & \mbox{for} \; n < 0 
\end{array} \right. ,
\label{bdef}
\end{eqnarray} 
and which obey the standard commutation relations.
The two-particle potential is denoted as $v(k)$, the Fermi velocity as $v_{\rm F}$, and momenta 
are given by $k_n=n 2 \pi /L$ (periodic boundary conditions) with system size $L$.   
For vanishing interaction $v(k)=0$ the $b_n^{(\dag)}$ are the eigenmode ladder operators 
and the boson dispersion is $\omega_0(k) = v_{\rm F} |k|$.   
We note that the interaction only couples the modes with fixed $|n|$. The Hamiltonian 
is thus a sum of commuting terms and the problem of finding the new eigenmodes 
factorizes. The eigenstates are product states (over $n \in {\mathbb N}$).   

By a Bogoliubov transform $H$ can straightforwardly be diagonalized 
\begin{eqnarray}
H  = \sum_{n \neq 0} \omega(k_n) \, \alpha_n^\dag \alpha_n^{}  +
E_{\rm gs}
\label{HTLdef}
\end{eqnarray} 
in terms of the eigenmodes
\begin{eqnarray} 
\alpha_n =  c(k_n) b_n - s(k_n) b_{-n}^\dag 
\label{bogo}
\end{eqnarray} 
with
\begin{eqnarray} 
&& s^2(k) = \frac{1}{2} \left[ \frac{1+ \hat v(k) /2 }{\sqrt{1+ \hat v(k)}} -1  \right] , \; 
c^2(k) = 1+ s^2(k) , \nonumber \\  && 
\omega(k) = v_{\rm F} \,  |k| \, \sqrt{1+\hat v(k)}  , 
\label{manydefs}
\end{eqnarray} 
and the dimensionless potential $\hat v(k) = v(k)/(\pi v_{\rm F})$. For small interactions 
\begin{equation}
s^2(k) = \hat v^2(k)/16 + {\mathcal O}(\hat v^3) .
\label{smallint}
\end{equation}
We assume that the Fourier transform $v(q)$ of the two-particle potential
is an even function which for $q>0$ decreases monotonically
on a characteristic 
scale $q_{\rm c}$. It can be expressed as a function of $q/q_{\rm c}$. 
In integrals over momenta we often 
substitute $q'=q/q_{\rm c}$  which implies $ \hat v(q) = \hat v(q_{\rm c} q')$, 
$c^2(q) = c^2(q_{\rm c} q')$, \ldots . 
We here focus on two-particle potentials with $\hat v(0) \geq 0$.  

The nondegenerate ground state of $H$ is given by the product state (over the mode index) 
of the vacua with respect to the eigenmodes $\left| \mbox{vac}(\alpha) \right>$ and 
\begin{eqnarray} 
E_{\rm gs} = - 2 \sum_{n>0}  \omega(k_n) \, s^2(k_n) 
\label{eee}
\end{eqnarray}  
is the ground-state energy. The excited states can be constructed by populating the 
bosonic states of the different eigenmodes $n$. Momentum conservation of the Hamiltonian
Eq.~(\ref{TLmodel}) implies that every excited state is at least doubly degenerate.  
Depending on the momentum dependence of the eigenmode dispersion $\omega(k)$ Eq.~(\ref{manydefs}) 
and thus the momentum dependence of the (dimensionless) two-particle potential $\hat v(k)$ 
further degeneracies might appear. Note that often $\omega(k)$ is linearized 
(in $k$) which leads to a vast number of degeneracies. However, all eigenstates are 
product states over the mode index $n \in {\mathbb N}$ and every factor is uniquely 
determined by the energy and the momentum contained in the given mode. Employing this one 
can generalize the proof of Sect.~\ref{sec:factor2} for time averages which is then 
based on an analog of Eq.~(\ref{weaker}) with the state being one of the factors of the 
many-body eigenstate. We refrain from going into details as even the direct applicability 
of the considerations of  Sect.~\ref{sec:factor2} for the TL model does not imply
that time-averaged and steady-state expectation values are equal (see the discussion 
of Sect.~\ref{smallquenchintro}).    
 
To be able to compute arbitrary fermionic expectation values one has to bosonize the  
field operator. We here focus on the right-moving particles with momentum space
ladder operators $c_{n,+}^{(\dag)}$ and field operator     
\begin{equation}
\label{field}
\psi_+^{\dagger}(x)= \frac{1}{\sqrt{L}} \sum_{n} e^{-ik_nx}  c_{n,+}^{\dagger} \;\; .
\end{equation} 
One can prove the operator identity \cite{Schoenhammer05,vonDelft98}
\begin{equation}
\label{bosonization1}
\psi^{\dagger}_+(x)= \frac{e^{- i x\pi/L}}{\sqrt{L}} 
e^{-i \Phi^{\dagger}(x)} U^{\dagger} e^{-i \Phi(x)} \;\; ,
\end{equation}
with
\begin{equation}
\label{bosonization2}
\Phi(x)=\frac{\pi}{L} N  x -i \sum_{n>0} e^{iq_nx} \left(
\frac{2\pi}{L q_n} \right)^{1/2}  b_n \;\; ,
\end{equation}
where $N$ denotes the particle number operator and $U^{\dagger}$ a unitary 
fermionic raising operator which commutes with the $ b_n^{(\dag)}$. It maps
the $N$-electron ground state to the $(N+1)$-electron one. Particle number
contributions do not matter for our considerations and will not be discussed
in detail.  We emphasize that the relation between the original fermionic degrees 
of freedom and the bosonic eigenmodes is highly nonlinear.

As the initial state we consider the ground state of the noninteracting 
Hamiltonian Eq.~(\ref{TLmodel}) with $v(k)=0$. It corresponds to the product state 
of the vacua with respect to the $b_n$,  $\left| E_{0}^{\rm i} \right> = 
\left| \mbox{vac}(b) \right>$, and the ground-state energy $E_0^{\rm i}$ 
vanishes. The excited states 
are constructed by filling the bosonic modes associated to the $b_n^{(\dag)}$. 
The time evolution is 
performed with the two-particle interaction switched on 
and $H_{\rm f}$ given by  Eq.~(\ref{TLmodel}). 
The (dimensionless) quench parameter is the amplitude of the two-particle potential
$\hat v(0)$ at vanishing momentum transfer.    

\subsection{The observables}

As `observables' we consider the kinetic energy 
\begin{align}
H_0 = \sum_{n \neq 0} v_{\rm F} |k_n| b_n^\dag b_n = H_{\rm i} ,
\label{obskin}
\end{align}
which is equivalent to the initial Hamiltonian, 
the operator content of the density-density correlation function of the right movers
\begin{align}
D(x)=\rho_+(x) \rho_+(0) = \frac{1}{L^2} \sum_{n,n' \neq 0} e^{i k_n x}
\rho_{n,+} \rho_{n',+} ,
\label{densdensdef}
\end{align}
as well as the one of the single-particle Green function of the right movers 
\begin{align}
G(x) = \psi_+^\dag(x) \psi_+(0)^{} .
\label{greendef}
\end{align}
We note that for arbitrary $x$, $D(x)$ and $G(x)$ do not commute with 
$H_{\rm i}=H_0$ and do thus not fall into the class of operators considered in
Refs.~\onlinecite{Moeckel08} and \onlinecite{Moeckel09}.

We explicitely show that $\tilde D(x)$ and $\tilde G(x)$ are `factor of two class' operators 
by computing their steady-state and interacting ground state expectation values. 

Employing Eq.~(\ref{bdef}) $D(x)$ can be written as
\begin{align}
D(x) = \frac{1}{2 \pi L} \sum_{l,l' > 0} & \sqrt{k_l k_{l'}} \left( e^{i k_l x} b_l b_{l'} + 
  e^{i k_l x} b_l b_{l'}^\dag \right. \nonumber \\
& \left. +  e^{-i k_l x} b_l^\dag b_{l'} +  e^{-i k_l x} b_l^\dag b_{l'}^\dag \right) .
\label{Drewritten}
\end{align}
To show that Eq.~(\ref{excondition}) defining the `thermalization class' holds for 
small $x$ we employ that $H_{\rm i}$ and  $H_{\rm f}$ both conserve the total momentum 
$\sim  \sum_{l \neq 0} k_l  b_l^\dag b_{l}$ and that $ b_l^{(\dag)}$ destroys (creates) a momentum
$k_l$. Both ensures that in the steady-state and thermal expectation values of 
$D(x)$ Eq.~(\ref{Drewritten}) only the second and the third 
term with $l=l'$ contribute. Furthermore, for $x=0$ the exponential terms in 
Eq.~(\ref{Drewritten}) are equal to 1 and $D(0)$ is proportional to the positive
momentum part of the kinetic energy $H_{\rm i}$. In the steady-state and thermal 
expectation values of the kinetic energy the positive and negative momentum parts 
of the sum contribute equally (by symmetry). 
Up to the prefactor $(2 \pi v_{\rm F} L)^{-1}$, these expectation 
values of $ D(0)$ are thus equal to the ones of  $H_0$. Thus the {\it local} operator 
$\tilde D(0)$ is element of the `thermalization class'. Comparing the steady-state and thermal 
expectation values of $D(0)$ we explicitely verify that it thermalizes to second order 
in the two-particle interaction.  

Employing Eqs.~(\ref{field}) to (\ref{bosonization2}) and the Baker-Campbell-Hausdorff formula 
$G(x)$ can be written as 
\begin{align}
G(x) = G_0(x)  & \exp\left\{ \sum_{l>0} \frac{e^{-ik_lx}-1}{\sqrt{l}} b_l^\dag \right\}
\nonumber \\ & \times \exp\left\{ \sum_{l'>0} \frac{1-e^{ik_{l'}x}}{\sqrt{l'}} b_{l'} \right\} ,
\label{Grewritten}
\end{align}      
with 
\begin{align}
G_0(x)= \frac{1}{L} \, \frac{e^{-i k_{\rm F} x}}{1-e^{i(2 \pi x/L+i0)}} .
\label{G_0}
\end{align}         
Following the same reasoning as for
$D(0)$ it is straight forward to show 
that 
\begin{align}
\frac{ \left< G(x) \right>_{\rm ss/th}}{G_0(x)}  = 1- x^2 \frac{2 \pi}{ L} \sum_{l>0}  
 k_l  \left< b_l^\dag b_{l} \right>_{\rm ss/th} + {\mathcal O}(x^4). 
\label{Gexpand}
\end{align}     
This shows that for small $x$, $G(x)$ is from the `thermalization class'.
Comparing the steady-state and thermal 
expectation values of $G(x)$ at small $x$ we explicitely verify that it 
thermalizes to second order.  

\begingroup
\squeezetable
\begin{table}
\begin{ruledtabular}
\begin{tabular}{cccc}
  observable  &  first order therm. & fact 2 cl. & therm. cl. \\ \hline
 $\tilde G(x)$  &  first order vanishes  & yes  & for $q_{\rm c} x \ll 1$ \\
  $\tilde D(x) $  & first order vanishes &  yes & for $q_{\rm c} x \ll 1$  \\
 $\tilde H_{\rm i}$ & first order vanishes & yes &  yes \\
  $ \rho(x) $  & yes &  no & no 
\end{tabular}
\end{ruledtabular}
\caption{Classification of the considered operators. \label{tableLL}}
\end{table}
\endgroup

We furthermore verify that Eq.~(\ref{21expl}) holds for  
\begin{align}
V=  \sum_{n >0 } & \left[  k_n  \frac{v(k_n)}{2 \pi \hat v(0)} 
\left( b_n^\dag b_n^{} + b_{-n}^\dag b_{-n}^{} \right) \right. \nonumber \\
& \left. + k_n   \frac{v(k_n)}{2 \pi \hat v(0)} 
\left( b_n^\dag b_{-n}^\dag + b_{-n}^{} b_{n}^{} \right)  \right]
\label{Vhier}
\end{align}
[see Eq.~(\ref{gVdef})]. With the results of Sect.~\ref{sec:secondordertherm} 
we can thus be sure that all `thermalization class' operators thermalize 
to second order in $\hat v(0)$. 

The classification of the considered operators is summarized in 
Table \ref{tableLL}.

\subsection{The time evolution and steady state}

Employing Eq.~(\ref{bosonization1}), the simple time dependence of the 
eigenmode ladder operators 
\begin{eqnarray} 
\alpha_n^{} (t) = e^{ -i \omega(k_n) t} \, \alpha_n^{} , \quad \alpha_n^{\dag} (t) 
= e^{ i \omega(k_n) t} \, \alpha_n^{\dag},
\label{modetime}
\end{eqnarray}
the ladder operator 
properties of the $b_n^{(\dag)}$ and $\alpha_n^{(\dag)}$, as well as the knowledge of 
the interacting and noninteracting ground states the time evolution of 
the expectation values of our observables out of the noninteracting ground state 
can be computed straightforwardly. After the thermodynamic limit was taken -- note 
that the kinetic energy is extensive and must first be divided by $L$ -- the limit $t \to \infty$ 
can be performed and the expectation values assume steady-state values.    

For the dimensionless kinetic energy per length we obtain
\begin{align}
e_{\rm ss} = \lim_{L \to \infty} \frac{\left< \tilde H_0 \right>_{\rm ss}}{v_{\rm F} q_{\rm c}^2 L} = 
\frac{2}{\pi}   \int_0^\infty dq  q  s^2(q_{\rm c} q) c^2(q_{\rm c} q) , 
\label{kinss}
\end{align}
where we have substituted $q/q_{\rm c} \to q$.  

The dimensionless steady-state expectation value of $\tilde D(x)$ is given by 
\begin{align}
q_{\rm c}^{-2} \left< \tilde D(x) \right>_{\rm ss} =  
\frac{1}{\pi^2} \int_0^\infty dq q  s^2(q_{\rm c} q) c^2(q_{\rm c} q)  \cos[q (q_{\rm c}x)] .
\label{densdensss}
\end{align}
If we expand the cosine function on the right-hand side for small $q_{\rm c} |x|$  
and keep only the leading term we obtain 
\begin{align}
q_{\rm c}^{-2} \left< \tilde D(x) \right>_{\rm ss}  = & 
\frac{1}{\pi^2}   \int_0^\infty dq q  s^2(q_{\rm c} q) c^2(q_{\rm c} q)  \nonumber
+ {\mathcal O}([q_{\rm c} x]^2) \nonumber
\\ 
= & \frac{1}{2 \pi} e_{\rm ss} + {\mathcal O}([q_{\rm c} x]^2) .
\label{densdenssssmallx}
\end{align} 
In the last step we used Eq.~(\ref{kinss}).
The relation between the local part of the steady-state expectation value of $\tilde D(x)$ and 
the kinetic energy is  in accordance with the discussion following Eq.~(\ref{Drewritten}). 

We finally consider the single-particle Green function $G(x)$. Its steady state expression was 
derived earlier; see e.g. Eq.~(33) of Ref.~\onlinecite{Rentrop12}. It can be written as   
\begin{align}
\left< \tilde G(x) \right>_{\rm ss}= G_0(x) \left[ e^{g_{\rm ss}(x)} -1 \right] ,
\label{Greenss}
\end{align}    
with 
\begin{align}
g_{\rm ss}(x) = 4 \int_0^\infty dq  s^2(q_{\rm c} q) c^2(q_{\rm c} q) \frac{\cos[q (q_{\rm c}x)] -1}{q}  .  
\label{smallgssdef}
\end{align}
For $q_{\rm c} |x| \ll 1$, $g_{\rm ss}(x)$  
further simplifies to
\begin{align}
g_{\rm ss}(x) & = - 2 (q_{\rm c}x)^2  \int_0^\infty dq  q  s^2(q_{\rm c} q) c^2(q_{\rm c} q) 
+ {\mathcal O}([q_{\rm c} x]^4)  \nonumber \\  
& = - \pi (q_{\rm c}x)^2  e_{\rm ss} 
+ {\mathcal O}([q_{\rm c} x]^4)  
\label{smallgsssmallx} 
\end{align} 
[for the second line see Eq.~(\ref{kinss})] and thus 
\begin{align}
\frac{\left<\tilde G(x)\right>_{\rm ss}}{G_0(x)} = - \pi (q_{\rm c}x)^2  e_{\rm ss} 
+ {\mathcal O}([q_{\rm c} x]^4)   ,
\label{Greensssmallx}
\end{align}   
in accordance with Eq.~(\ref{Gexpand}).

The steady-state expectation values can also be obtained using a GGE.\cite{Cazalilla06}
As for the quench in the tight-binding chain we, however, preferred to 
compute the full time evolution to verify that all the observables 
of interest become stationary. 

\subsection{The thermal expectation values}

We next compute the thermal expectation values.  

The effective temperature corresponding 
to the energy quenched into the system is determined by 
\begin{align} 
& \lim_{L \to \infty} \frac{\left<  \mbox{vac}(b) \right| H_{\rm f} \left|  \mbox{vac}(b) \right>}{L}  
\overset{!}{=}  \lim_{L \to \infty}  \frac{1}{L Z_{\rm f}} \mbox{Tr} \left( e^{-H_{\rm f}/T_{\rm f}} H_{\rm f} \right)
\nonumber \\ 
 \Leftrightarrow & 0 \overset{!}{=}  \int_0^\infty dk \omega(k)  \left[
n\left( \frac{\omega(k)}{T_{\rm f}} \right) - s^2(k) \right],   
\label{T_effdet} 
\end{align}
where $n(x)= \left[  e^{x} -1  \right]^{-1}$ is the Bose function. 
We note that for the phonon-like bosons $\alpha_n^{(\dag)}$ the chemical potential vanishes. 
Equation (\ref{T_effdet}) can straightforwardly be solved numerically. For small quenches 
$|\hat v(0) | \ll 1$ we can make analytical progress using Eq.~(\ref{smallint})   
and obtain
\begin{align}
\left( \frac{T_{\rm f}}{v_{\rm F} q_{\rm c}} \right)^2 =  \frac{3}{8 \pi^2}  
\int_0^\infty dq q \hat v^2(q_{\rm c} q) +  {\mathcal O}\left(\hat v^3(0)\right) .
\label{T_effres} 
\end{align} 
This is consistent with Eq.~(\ref{Texpansecond}). 

The thermal expectation values of our observables can be computed employing the same 
ideas as used for the time evolution. They are given by 
\begin{align}
e_{\rm th} = & \frac{1}{\pi} \int_0^\infty  dq  q  
\nonumber \\ & \times  \left[ s^2(q_{\rm c} q) + \left\{ 1 + 2 s^2(q_{\rm c} q) \right\}  
n\left(\frac{\omega(q_{\rm c} q)}{T_{\rm f}}\right) \right]
\label{kinth}
\end{align}
for the kinetic energy,
\begin{align}
q_{\rm c}^{-2} & \left< \tilde D(x) \right>_{\rm th} = 
\frac{1}{2\pi^2} \int_0^\infty dq q \cos[q (q_{\rm c}x)] \nonumber \\ & \times 
\left[ s^2(q_{\rm c} q) + \left\{ 1 + 2 s^2(q_{\rm c} q) \right\}  
n\left(\frac{\omega(q_{\rm c} q)}{T_{\rm f}}\right) \right] . 
\label{Dth}
\end{align}
for the density-density-correlation function, and
\begin{align}
\left< \tilde G(x) \right>_{\rm th}= G_0(x) \left[ e^{g_{\rm th}(x)} -1 \right] ,
\label{Greenth}
\end{align}   
with
\begin{align}
g_{\rm th}(x) & =  2 \int_0^\infty dq \frac{\cos[q (q_{\rm c}x)] -1}{q}  \nonumber \\
&\times 
  \left[ s^2(q_{\rm c} q) + \left\{ 1 + 2 s^2(q_{\rm c} q) \right\}  
n\left(\frac{\omega(q_{\rm c} q)}{T_{\rm f}}\right) \right]  
\label{smallgthdef}
\end{align}
for the single-particle Green function. Remarkably, in all expressions the interaction and temperature
enter via the same factor given by the square brackets.  

\subsection{Comparison}

We are now in a position to explicitly confirm Eq.~(\ref{moeckelallg}) 
(with $\overline{\left< \ldots \right>}$ replaced by $\left< \ldots \right>_{\rm ss}$)
for all the considered 
observables. Expanding the steady-state expectation values Eqs.~(\ref{kinss}), (\ref{densdensss}), 
and (\ref{Greenss}) [with Eq.~(\ref{smallgssdef}) inserted] for $|\hat v(0)| \ll 1$ leads to    
\begin{align}
e_{\rm ss} = 
\frac{1}{8 \pi}   \int_0^\infty dq  q  \hat v^2(q_{\rm c}q) + {\mathcal O}\left(\hat v^3(0)\right) , 
\label{kinsssmall}
\end{align}
\begin{align}
q_{\rm c}^{-2} \left< \tilde D(x) \right>_{\rm ss} = &  
\frac{1}{16 \pi^2} \int_0^\infty dq q  \hat v^2(q_{\rm c} q)   \cos[q (q_{\rm c}x)] \nonumber \\ 
& + {\mathcal O}\left(\hat v^3(0)\right) ,
\label{densdenssssmall}
\end{align}
and 
\begin{align}
\frac{\left< \tilde G(x) \right>_{\rm ss}}{G_0(x)} = &
\frac{1}{4} \int_0^\infty dq  \hat v^2(q_{\rm c} q)  \frac{\cos[q (q_{\rm c}x)] -1}{q} \nonumber \\  & + {\mathcal O}\left(\hat v^3(0)\right) , 
\label{Greensssmall}
\end{align}    
where we used Eq.~(\ref{smallint}). Setting $T_{\rm f}=0$ and thus $n(\omega(q_{\rm c} q)/T_{\rm f}) \to 0$  
in Eqs.~(\ref{kinth}), (\ref{Dth}), and  (\ref{smallgthdef}) we obtain   
the ground-state expectation values of our observables with respect to $H_{\rm f}$. If we 
expand the resulting expressions for $|\hat v(0)| \ll 1$ employing Eq.~(\ref{smallint}) we recover 
Eqs.~(\ref{kinsssmall}), (\ref{densdenssssmall}), and  (\ref{Greensssmall})  up to a factor $1/2$ which
confirms  Eq.~(\ref{moeckelallg}); all the considered observables are from the `factor of two class'.      

We next explicitly investigate thermalization to order $\hat v^2(0)$. We start out 
with the thermal kinetic energy Eq.~(\ref{kinth}). For $T_{\rm f}>0$ also the second 
term in the square brackets associated to thermal excitations contributes. For small quenches 
we can use that $T_{\rm f}$ is of order $\hat v(0)$ [see Eq.~(\ref{T_effres})] and obtain 
\begin{align}
e_{\rm th} = & \frac{1}{\pi} \int_0^\infty  dq  q \Big[
\underbrace{\frac{1}{16} \hat v^2(q)}_{T_{\rm f}=0-\mbox{\scriptsize contr.}}
+ \underbrace{n\left(\frac{\omega_0(q_{\rm c} q)}{T_{\rm f}}\right)}_{T_{\rm f}>0-\mbox{\scriptsize contr.}} \Big] + {\mathcal O}\left(\hat v^3(0)\right) \nonumber \\
 = & \frac{1}{\pi} \int_0^\infty \!\!\!\!\!  dq  \Big[
\underbrace{\frac{q}{16} \hat v^2(q)}_{T_{\rm f}=0-\mbox{\scriptsize contr.}}
+ \underbrace{\left(\frac{T_{\rm f}}{v_{\rm F} q_{\rm c}}\right)^2 \frac{q}{e^{q}-1}}_{T_{\rm f}>0-\mbox{\scriptsize contr.}} 
\Big] + {\mathcal O}\left(\hat v^3(0)\right)  \nonumber \\
= & \frac{1}{\pi} \int_0^\infty   \!\!\!\!\!   dq  \Big[
\underbrace{\frac{q}{16} \hat v^2(q)}_{T_{\rm f}=0-\mbox{\scriptsize contr.}}
+ \underbrace{\frac{q}{16} \hat v^2(q)}_{T_{\rm f}>0-\mbox{\scriptsize contr.}} 
\Big] + {\mathcal O}\left(\hat v^3(0)\right) .
\label{kinthexp}
\end{align}    
In the last step we used   Eq.~(\ref{T_effres}) and $\int_0^\infty dq q/(e^{q}-1) =\pi^2/6 $.  
The ground-state contribution and the one from the thermal excitations at $T_{\rm f}$ -- both 
being equal -- add up to give the 
steady-state expectation value Eq.~(\ref{kinsssmall}) up to corrections of order $\hat v^3(0)$.  
This confirms thermalization to quadratic order and provides yet another example
for our explanation of the factor of 2 in Eq.~(\ref{moeckelallg}) for observables which are 
simultaneously from the `factor of two class' and the `thermalization class'.

After expanding the cosine appearing in Eqs.~(\ref{Dth}) and (\ref{smallgthdef}) to 
lowest nonvanishing order the same steps can be performed as the remaining
integrals are exactly of the form which appeared in the analysis of the kinetic energy. 
Thus $D(x)$ thermalizes to order $\hat v^2(0)$ up to corrections of order $(q_{\rm c}x)^2$ 
and $G(x)$ up to corrections of order $(q_{\rm c}x)^4$.  In other words, the {\it local} 
parts of $D(x)$ and $G(x)$ thermalize for small quenches. The factor 
of two  of Eq.~(\ref{moeckelallg}) can again be traced back to the two equal contributions 
from the ground state and the thermal excitations.
 
This is in full agreement with
the general results of Sect.~\ref{sec:secondordertherm}. 
The operator content of the first term of $V$  Eq.~(\ref{Vhier}) is identical to that 
of the kinetic energy. For the latter we have just shown that the steady-state and 
thermal expectation values are identical up to corrections of order $\hat v^3(0)$. 
Thus the same holds for this contribution to $V$. To show that also the second term
in $V$ fulfills Eq.~(\ref{21expl}) does also not require any additional computations.
When computing thermal expectation values of products of two $b$ ladder operators 
temperature always enters via the Bose function. As we have just seen when going from
the first to the second line in Eq.~(\ref{kinthexp}) $q$ times the Bose function always 
contributes a factor $T_{\rm f}^2 \sim \hat v^2(0)$. The $T_{\rm f}=0$  and $T_{\rm f}>0$ 
expectation values thus agree up to corrections of order $ \hat v^2(0)$ and  
Eq.~(\ref{21expl}) holds for $V$ of Eq.~(\ref{Vhier}). 

\subsection{The density in the semi-infinite Tomonaga-Luttinger model}

The observables we considered so far for the TL model were all from 
the `factor of two class' and, if local, in addition from the `thermalization class'.
To provide an example of an observable which does not fall into the `factor of two class',
we investigate the density $\rho_+(x)= \psi_+^\dag(x) \psi_+(x)$.
For a translationally invariant system its expectation value  is  $x$-independent and turns 
out to be time-independent after the interaction quench. 
Neither is true if we 
consider the TL model with open boundary conditions.\cite{Fabrizio95} 
The quench dynamics of the TL model in the presence of open boundaries was studied earlier.\cite{Kennes13} 
The exact computation of 
the quench dynamics of this model requires so-called open boundary bosonization.\cite{Fabrizio95,Meden00,Grap09} We 
here refrain from giving details and only report on the results. The steps required to obtain 
these are very similar to the ones taken above for the quench dynamics of the translationally 
invariant TL model.

In the long-time limit the density becomes stationary and its steady-state expectation 
value $d_{\rm ss}(x)$ is given by
\begin{align}
\frac{d_{\rm ss}(x)}{d_0(x)} =  \exp\Biggl\{ & \int_0^\infty dq  2 s(q_{\rm c} q) c(q_{\rm c} q)
 \left[ s(q_{\rm c} q) + c(q_{\rm c} q) \right]^2  \nonumber \\ &  \times 
 \frac{\cos[q (q_{\rm c}x)] -1}{q} \Biggr\} ,
\label{densss}
\end{align}
where $d_0(x)$ denotes the density of the open system at vanishing two-particle interaction.  
The thermal expectation value is 
\begin{align}
\frac{d_{\rm th}(x)}{d_0(x)} = \exp\Biggl\{  \int_0^\infty dq  2 & \Biggl[ s^2(q_{\rm c} q) +  
s(q_{\rm c} q) c(q_{\rm c} q) \nonumber \\ & +
\left\{ s(q_{\rm c} q) + c(q_{\rm c} q) \right\}^2 n\left(\frac{\omega(q_{\rm c} q)}{T_{\rm f}}\right)
\Biggr] \nonumber \\ &  \times 
 \frac{\cos[q (q_{\rm c}x)] -1}{q} \Biggr\}. 
\label{densth}
\end{align}
Via the term $s(q_{\rm c} q) c(q_{\rm c} q) $ the interaction now enters to linear order in $\hat v(0)$;
$\rho_+(x)$ of the semi-infinite TL model is not from the  `factor of two class'. 
To order $\hat v(0)$  the argument of the exponential function in $d_{\rm ss}(x)$ and $d_{\rm th}(x)$   
is equal to $- \hat v(q_{\rm c}q) \{ \cos[q (q_{\rm c}x)] -1 \}/(2 q) $ and the 
density thermalizes to order $\hat v(0)$.  We emphasize that this holds independently of the position
$x$ away from the open boundary; there is no requirement of locality for first order thermalization. 
Furthermore, to leading order $T_{\rm f}$ in Eq.~(\ref{densth}) 
can be set to zero and Eq.~(\ref{firstorder}) 
with $\overline{\left< \ldots \right>}$ replaced by $\left< \ldots \right>_{\rm ss}$
holds.

\section{Interaction quenches for spinless lattice fermions}
\label{sec:lattice}

\subsection{The Hamiltonians and DMRG}

Finally, we study interacting spinless fermions on a lattice of size $M$ described by the Hamiltonian
\begin{align}
H = &  - \frac 12 \sum_{j=1}^{M} \left( c_{j+1}^\dag c_j + \mbox{H.c.} \right) \nonumber \\ 
& + U \sum_{j=1}^M 
\left(n_j-\frac 12\right)\left(n_{j+1}-\frac 12\right),
\label{HIFnN}
\end{align}
where $n_j= c_{j}^\dag c_j$ is the density. The nearest-neighbor density-density-type 
two-particle interaction has the (dimensionless) strength $U$. We focus 
on repulsive interactions $U>0$ and half-filling of the band. 
Note that in contrast to the Hamiltonian of Sect.~\ref{sec:tightbinding} where the amplitude
of the nearest-neighbor hopping $J$ was set to $1$ it will turn out to be more
convenient to  take $J=1/2$ here. 
With this choice the bandwidth is two. The model can be mapped onto the 
XXZ-Heisenberg spin chain
\begin{align}
H =   \sum_{j=1}^{M} \left( S^x_{j+1} S^x_j + S^y_j S^y_{j+1} \right) + U \sum_{j=1}^M 
S^z_jS^z_{j+1},
\label{HIFnNSpin}
\end{align} 
using a Jordan-Wigner transform. The $S_j^{\zeta}$, with $\zeta=x,y,z$, denote spin-1/2 
operators. This simple form justifies a posteriori the choice $J=1/2$.
The spin Hamiltonian is convenient when using DMRG. 

The XXZ-Heisenberg spin chain is Bethe 
ansatz solvable. Therefore, many of its equilibrium properties are known 
analytically (see e.g. Refs.~\onlinecite{Giamarchi03} and \onlinecite{Schoenhammer05}). 
The system is gapless for $0\leq U<1$, while for $U>1$ 
a gap opens. In the latter regime long-ranged anti-ferromagnetic ordering emerges -- 
charge-density-wave order in the language of the fermions. The former displays 
(metallic) Luttinger liquid behavior. 
The model is characterized by an extensive set of 
(quasi-) local integrals of motion and thus constitutes a prototypical model of the class 
(iii) defined in the Introduction. 

The implications of this set of conserved quantities on the relaxation dynamics 
was heavily debated.\cite{Gogolin15}  
We stress that our general considerations of Sect.~\ref{sec:general} 
do not make use of integrals of motion and our predictions are 
thus valid in the presence as well as in the absence of such a set. To emphasize this we in addition 
study a model in which the nearest-neighbor interaction is supplemented by a  
next-nearest-neighbor one with (dimensionless) strength $W\geq 0$
\begin{align}
H =&  \sum_{j=1}^{M} \left( S^x_{j+1} S^x_j + S^y_j S^y_{j+1} \right) + U \sum_{j=1}^M 
S^z_jS^z_{j+1}\notag\\
&+ W \sum_{j=1}^M 
S^z_jS^z_{j+2}.
\label{HIFnnNSpin}
\end{align} 
For such a system a Bethe ansatz solution is not known.
It is generally believed that the model is not solvable employing this method.
The model thus represents the most generic class (iv). The 
equilibrium phase diagram 
including the next-nearest-neighbor interaction is more complicated, but displays an extended 
gapless phase at small $U$ and small $W$. 

Besides the degeneracies associated to translation invariance at least for $U=W=0$ 
further `accidental' degeneracies can be found.\cite{Deguchi01,Fagotti14a} 
The same considerations as in 
Sect.~\ref{sec:tightbinding} hold and Eq.~(\ref{weaker}) cannot be exploited directly.    

As the initial state we consider the nondegenerate ground state 
$\left|E^{\rm i}_0\right>$ for $U=W=0$ (odd number of particles) and perform 
the time evolution with respect to finite $U$ and possibly $W$. Thus for small quenches
the system will remain in its gapless phase, which is at the center of our interest.
For completeness we will also show results for gapped systems (after the quench) to 
illustrate how the perturbatively motivated analytical insights become invalid at larger 
interactions.

As the Hamiltonians introduced above include two-particle interactions which are 
difficult to treat analytically, we rely on the numerically exact DMRG in the following. 
The accuracy of the DMRG is (in practice) controlled by a numerical parameter called the 
bond dimension $\chi$. Increasing $\chi$ and with it the numerical effort, one can 
achieve converged results. In the following $\chi$ is always chosen such that no changes 
of the results can be observed on the scales of the respective plots if it is further 
increased (`numerically exact results'). 

We are left with performing three tasks: (a) 
preparing the ground state of the noninteracting ($U=0=W$) system, (ii) subsequently 
performing the time evolution with respect to finite $U$ and possibly $W$ and (iii) 
calculating the finite temperature canonical ensemble with respect to finite $U$ and 
possibly $W$ (which is independent of the previous two tasks). 

We implement our DMRG algorithm using the language of matrix product states 
(MPS) and so-called infinite boundary conditions, which can 
be coded very elegantly. More importantly, one obtains results directly in the 
thermodynamic limit $M\to \infty$\cite{Vidal07,Orus08} and does not need 
to perform a finite-size scaling analysis.\cite{Karrasch12a} 
As the algorithms are well 
documented\cite{Schollwoeck11} we will skip the technical details and provide 
only an overview. The ground state is found via an iterative procedure. One starts 
with a random MPS and repetitively applies $e^{-H_{\rm i}\Delta\tau}$, normalizing the MPS after 
each step. To apply $e^{-H_{\rm i}\Delta\tau}$ we choose a second order Trotter decomposition. 
We start with larger values of $\Delta\tau$ and successively lower it until the MPS converges 
to $\left|E^{\rm i}_0\right>$. We aim at a relative accuracy of the total energy per lattice 
site of $10^{-10}$; after this is reached we stop the algorithm. It is straightforward to 
benchmark the outcome of this procedure, as the ground state preparation is done for the 
noninteracting system $U=0=W$. Next we perform a real-time evolution of this MPS by 
repetitively applying $e^{-iH_{\rm f}\Delta t}$, 
where we use a fourth order Suzuki-Trotter decomposition and choose $\Delta t$ small enough such 
that the error arising from this decomposition is negligible. To determine the finite temperature 
canonical ensemble we use purification at $\beta=1/T=0$ effectively rewriting the ensemble as an 
MPS in enlarged physical space.\cite{Verstraete04} Afterwards one can `cool down'
the MPS by repetitively applying $e^{-H_{\rm f}\Delta\beta}$ (and normalizing) in an 
appropriately Trotter decomposed fashion.

\subsection{The observables}

As in Sects.~\ref{sec:tightbinding} and \ref{sec:TLmodel} we consider 
several observables. We study the spin-spin correlations of the $z$-direction 
\begin{equation}
 C^{zz}_{r}= S^z_j S^z_{j+r},
\end{equation}
which due to translation invariance are $j$-independent. 
As explicitly discussed below (see Figs.~\ref{fig:DMRG_relax_small} and \ref{fig:DMRG_relax}) 
$\left< E_0^{\rm i} \right|  C^{zz}_{r}(t) \left| E_0^{\rm i} \right>$ 
converges towards a steady-state value for all $r$ and all parameter sets studied. 
The expectation values $\left\langle C^{zz}_{r} \right\rangle_{\rm ss/th}$ turn out to have a leading 
linear order dependence on the quench amplitude and $C^{zz}_{r}$ is  neither from the 
`factor of two class' nor the `thermalization class'. 
As shown in the lower panels of Figs.~\ref{fig:comp_nI} (lower three) and \ref{fig:comp_nnI} (lower two)
the numerical results for the steady-state and ground state (with respect to $H_{\rm f}$) expectation 
values at sufficiently small $U$ are fully consistent with Eq.~(\ref{firstorder}) 
where $\overline{\left< \ldots \right>}$ is replaced by $\left< \ldots \right>_{\rm ss}$ 
(compare the crosses and the solid lines).  We thus expect Eq.~\eqref{eq:firstordertherm} 
to hold for $\left\langle C^{zz}_{r} \right\rangle_{\rm ss/th}$ 
indicating thermalization to linear order. 

We furthermore compute the kinetic energy per lattice site 
\begin{equation}
\tilde h_{\rm kin}= \frac 1M  \left(H_{\rm{i}}- 
\left\langle E_0^{\rm i}\right|H_{\rm{i}}\left|E_0^{\rm i}\right\rangle\right),
\end{equation}
where we have subtracted the value of the kinetic energy before the quench to eliminate an irrelevant 
constant. It takes a steady-state value; see Figs.~\ref{fig:DMRG_relax_small} and \ref{fig:DMRG_relax}. 
The numerical results for the  steady-state expectation value of the kinetic energy 
are consistent with an order $g^2$ dependence; see the crosses in the 
upper panel of  Figs.~\ref{fig:comp_nI} and \ref{fig:comp_nnI}. A comparison of the crosses 
with the solid lines which show twice the ground state (with respect to $H_{\rm f}$) 
expectation values indicate that  Eq.~(\ref{moeckelallg}) with $\overline{\left< \ldots \right>}$ 
replaced by $\left< \ldots \right>_{\rm ss}$ holds; the kinetic energy is a `factor of two class' 
operator. It is per definition from the `thermalization class' and we expect thermalization to 
second order.

\begingroup
\squeezetable
\begin{table}
\begin{ruledtabular}
\begin{tabular}{cccc}
  observable  &  first order therm. & fact 2 cl. & therm. cl. \\ \hline
 $ C^{zz}_{r}$  &  yes  & no  & no  \\
  $\tilde h_{\rm kin} $  &  first order vanishes &  yes & yes  \\
 $\tilde B_r $ & first order vanishes & yes &  for $r=1$ 
\end{tabular}
\end{ruledtabular}
\caption{Classification of the considered operators. \label{tableXXZ}}
\end{table}
\endgroup

Finally, we introduce observables which are of the `factor of two class', but not necessarily 
of the `thermalization class'. A simple example of such are the fermionic bond operators
\begin{equation}
\tilde B_r= c_j^\dagger c_{j+r}+{\rm H.c.}- \left\langle E_0^{\rm i}\right|c_j^\dagger c_{j+r}
+{\rm H.c.}\left|E_0^{\rm i}\right\rangle,
\end{equation} 
which take steady-state values (not shown). The system remains translationally 
invariant by one lattice site even after the quench. The expectation values (during the time 
evolution, in the steady state as well as in the thermal state) of the operator
will thus be independent of $j$. This explains why we suppressed the index $j$ in the definition
of $\tilde B_r$. The sum of $\tilde B_r$ over $j$ can be expressed in terms of the 
noninteracting eigenmode number operators $c_k^\dag c_k$. Those in turn commute with $H_{\rm i}$ 
and the condition (c) of Refs. \onlinecite{Moeckel08} and \onlinecite{Moeckel09} to prove 
Eq.~(\ref{moeckel}) is fulfilled; our generalization is not required. The combination of 
translational and particle-hole symmetry (at half-filling) ensures that 
$\left\langle \tilde  B_r\right\rangle_{\rm ss}
=0=\left\langle \tilde B_r\right\rangle_{\rm th}$ for even $r$. Thus we focus on $r$ being odd. 
As exemplified in Fig.~\ref{fig:fac2_comp_nI} all $ \tilde B_r$ are element of the 
`factor of two class' (compare the crosses and solid lines).
However, only $\tilde B_1$ can be related to the kinetic  energy due to spatial translation 
invariance and is thus also of the `thermalization class'. Thus Eq.~(\ref{moeckelallg}) 
with $\overline{\left< \ldots \right>}$ replaced by $\left< \ldots \right>_{\rm ss}$ holds for 
all  $\tilde B_r$ but thermalization to second order can only be expected for $\tilde B_1$.

The classification of the considered operators is summarized in 
Table \ref{tableXXZ}.

\subsection{Time evolution towards the steady state}
\label{sec:timeevol}

First, we benchmark the  time scales accessible within our numerical DMRG approach 
at given computational resources and show that on these all observables of interest take 
steady-state values. To reach a designated accuracy in DMRG simulations the bond dimension 
$\chi$ must be increased for increasing entanglement of the system. As entanglement 
generically grows with time one can only reach a finite time scale (for given numerical 
resources).

We show the dynamics resulting out of different quench protocols in 
Figs.~\ref{fig:DMRG_relax_small} and \ref{fig:DMRG_relax} for two 
different small quench amplitudes.
 The time evolution of the observables shows an initial transient behavior on times of the order of 
the inverse bandwidth. Within this transient regime the observables exhibit damped oscillatory behavior, 
where the amplitude almost perfectly dies out (on the scale of the plots) after times of the order of 
ten times the inverse bandwidth. Generically, the lower the value of $U$  and $W$ the larger is 
the time scale which can be reached at given resources (compare Figs.~\ref{fig:DMRG_relax_small} 
and \ref{fig:DMRG_relax}). 
This is very reasonable as larger 
quenches result in a larger entropy production and thus a larger entanglement growth. In any 
case one can access time scales of the order of a $100$ or $10$ times the inverse bandwidth for 
smaller and larger quenches, respectively. This is sufficient to extract what appears to be 
the steady value to high precision for all our observables and all parameter sets considered. 
This demonstrates the usefulness of the DMRG for the analysis we have in mind. Note that 
there is no notable difference in the time evolution 
for the cases of vanishing and finite $W$, that is for systems with an extensive set of (quasi-) local 
integrals of motion and without this. 
We
emphasize, that we studied the time evolution for many
more parameter sets and always found a behavior similar to the one 
shown in Figs.~\ref{fig:DMRG_relax_small} and \ref{fig:DMRG_relax}.

\begin{figure}[t]
\centering
\includegraphics[width=\columnwidth]{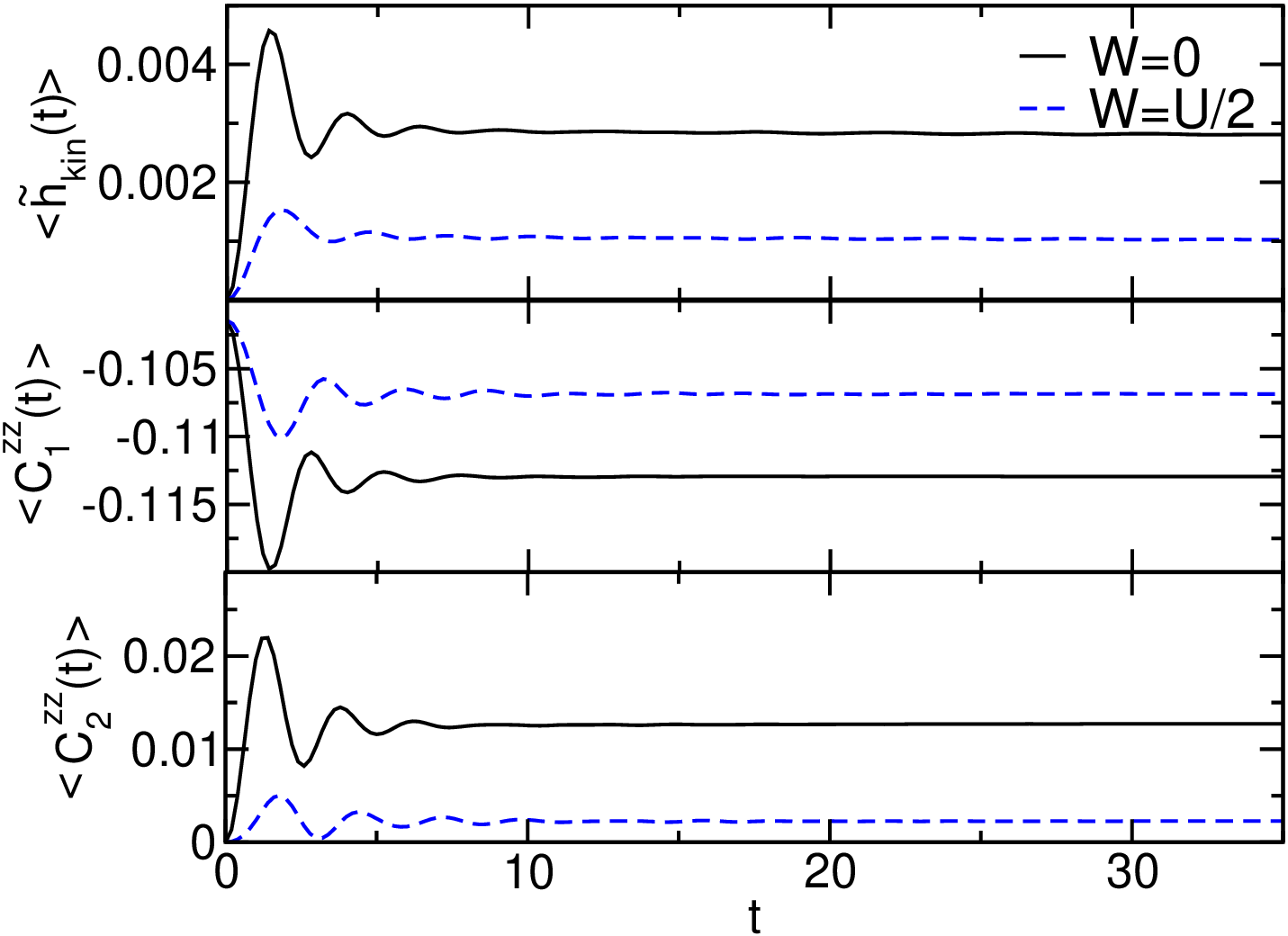}
\caption{Time evolution of $\left< O(t) \right> = 
\left< E_{0}^{\rm i} \right| O(t) \left| E_{0}^{\rm i} \right> $ for 
different observables $O$ after an interaction quench 
with amplitude $U=0.25$, $W=0$ and $U=0.25$, $W=0.125$ obtained by DMRG. The initial state is the 
noninteracting ground state. For all observables the results become stationary for the times reachable.}
\label{fig:DMRG_relax_small}
\end{figure} 

\begin{figure}[t]
\centering
\includegraphics[width=\columnwidth]{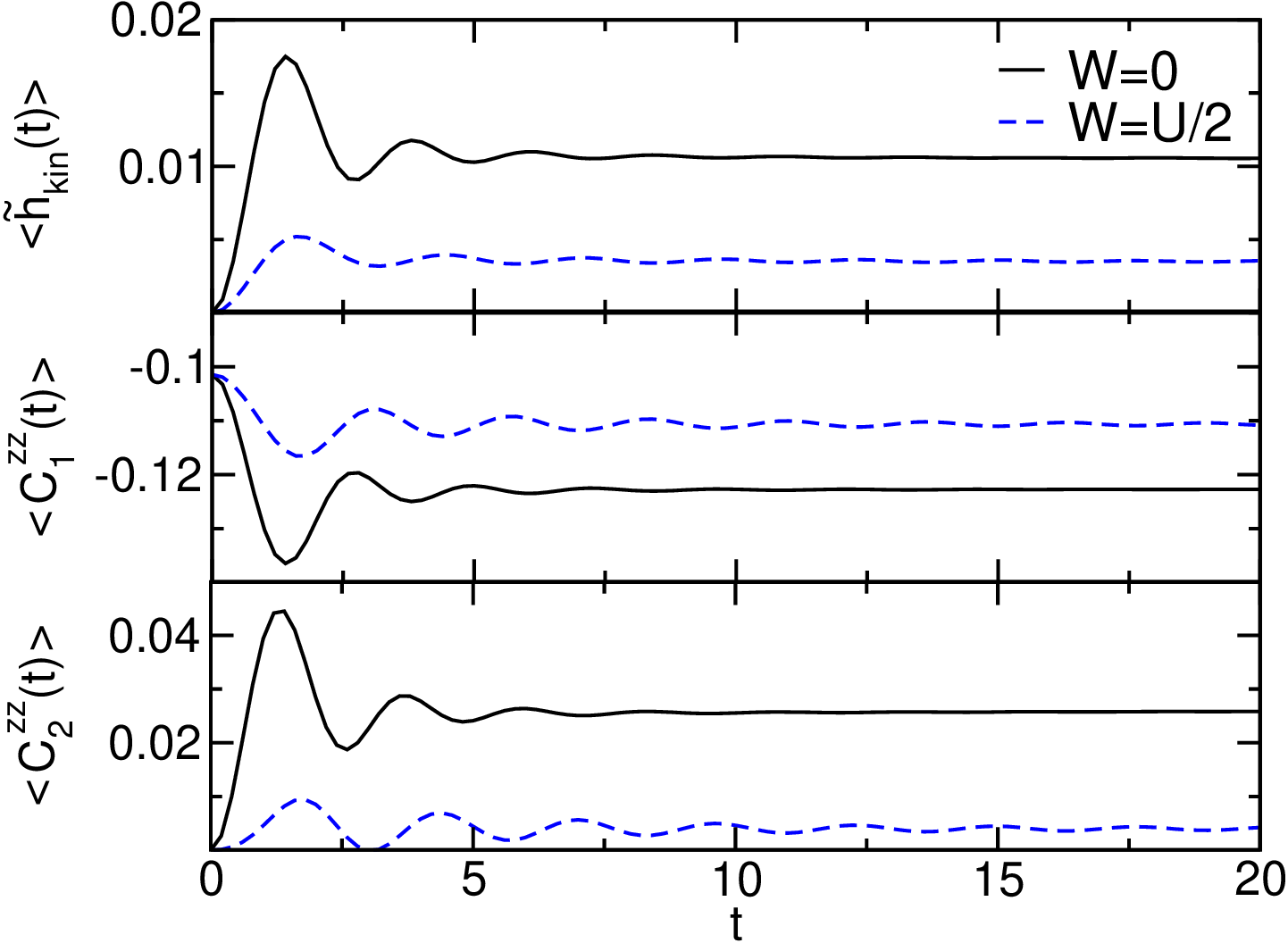}
\caption{The same as in Fig.~\ref{fig:DMRG_relax_small} but for  $U=0.5$, $W=0$ and $U=0.5$, $W=0.25$. The larger
the interaction quench the stronger the entanglement grows with time and the shorter the time scales reachable 
for given computational resources (compare to Fig.~\ref{fig:DMRG_relax_small}).} 
\label{fig:DMRG_relax}
\end{figure}

In the following, the steady state value of an observable is 
approximated by its value at the largest time reached within our simulation. Obviously 
purely based on the numerics we cannot rule out that at much larger time scales another relaxation 
mechanism sets in leading towards a different steady-state value. However, we cannot observe an 
indication of this in any of our data sets (for $W=0$ and $W>0$). In fact, the agreement of the 
steady-state value read off from the numerical data with the 
thermal expectation value according to our general considerations of Sects.~\ref{sec:firstordertherm} 
and \ref{sec:secondordertherm} we discuss next  gives us confidence that we have reached the 
ultimate steady state value. We emphasize that the observed single step relaxation 
of {\it local} observables is in accordance with the prethermalization conjecture.\cite{Berges04}

\subsection{Steady state for the model with nearest-neighbor interaction}
\label{sec:nearest}

We next compare the steady-state expectation value of our observables obtained after an 
interaction quench to their thermal counterparts. The temperature $T_{\rm f}$ of the 
thermal ensemble 
is computed using DMRG and Eq.~\eqref{Teffdef} by an iterative procedure.  
The inset of Fig.~\ref{fig:comp_nI} shows the effective temperature determined this way 
in dependence of $U$ after the quench in comparison to our order $U$ prediction 
Eq.~\eqref{Texpansecond}.

\begin{figure}[t]
\centering
\includegraphics[width=\columnwidth]{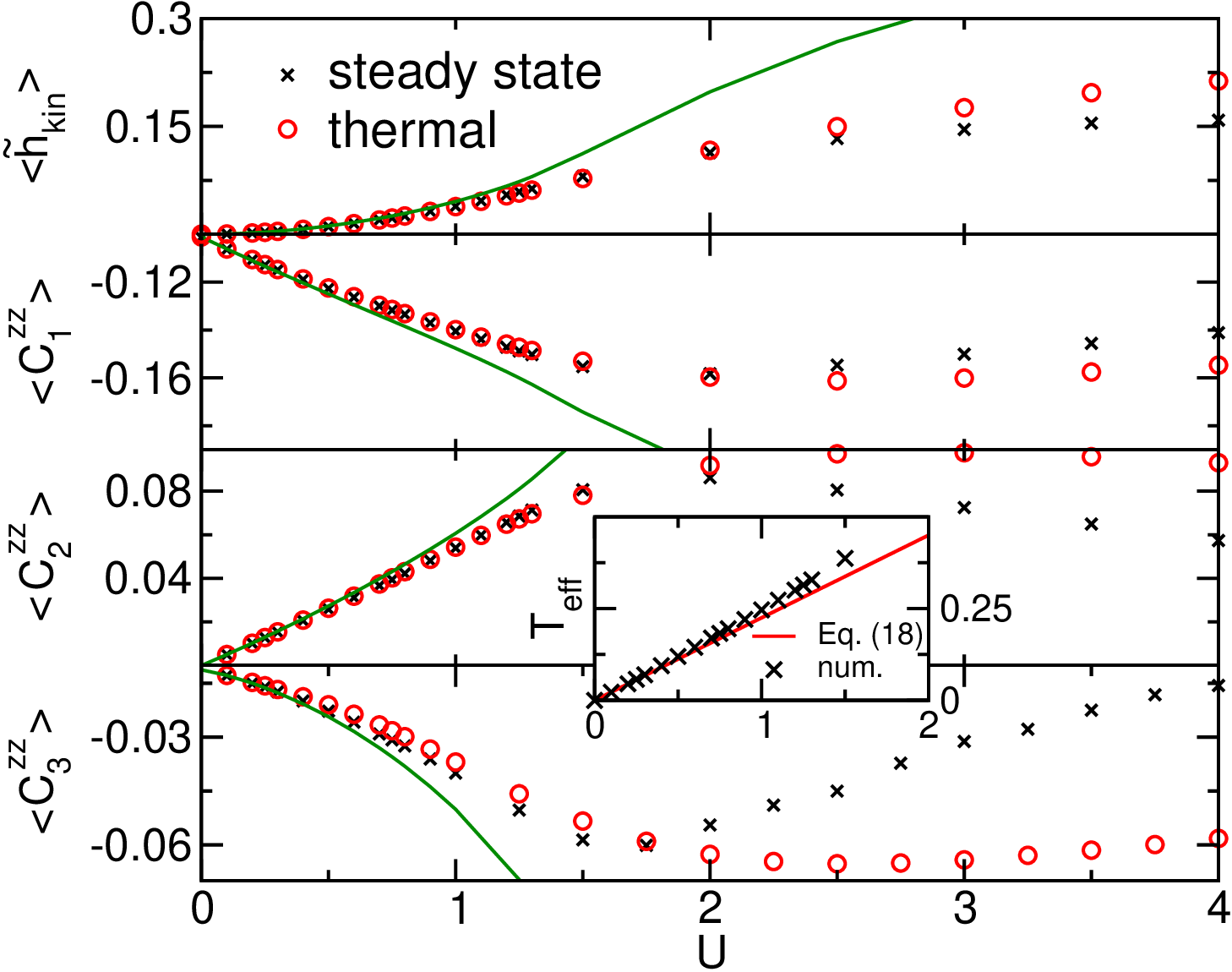}
\caption{Comparison of the steady-state (crosses) and thermal (circles) expectation value for 
different observables as a function of the quench amplitude $U$. The kinetic energy (upper panel) is a 
`factor of two class' as well as a `thermalization class' operator. Accordingly its leading 
term is ${\mathcal O}(U^2)$ with agreeing  prefactors of the steady-state and thermal values. 
The solid line shows twice the ground state (with respect to $H_{\rm f}$) expectation value 
[see Eq.~(\ref{moeckelallg}) with with $\overline{\left< \ldots \right>}$ 
replaced by $\left< \ldots \right>_{\rm ss}$]. 
The spin-spin correlations (lower three panels) are neither from the `factor of two class' nor the 
`thermalization class' and show thermalization to order $U$. The more local the observable, 
that is, the smaller $r$, the longer the linear order term prevails. 
The solid lines show the ground state (with respect to $H_{\rm f}$) expectation values
[see  Eq.~(\ref{firstorder}) with with $\overline{\left< \ldots \right>}$ 
replaced by $\left< \ldots \right>_{\rm ss}$].
The inset shows the $U$-dependence of the effective 
temperature $T_{\rm f}$ (over the entire gapless phase) obtained numerically as well as from the 
order $U$ expression Eq.~(\ref{Texpansecond}).}
\label{fig:comp_nI}
\end{figure} 

First, we focus  on $W=0$ [model class (iii)] and consider the kinetic energy $\tilde h_{\rm kin}$ as well 
as the spin-spin correlations $C_r^{zz}$ with $r\in \{1,2,3\}$ as typical examples of {\it local} observables. 
As argued above, the kinetic energy is of the `factor of two class' (compare crosses and line in the 
upper panel of  Fig.~\ref{fig:comp_nI}) and by definition from 
the `thermalization class', while the spin-spin correlator is neither element of the former nor the latter.
The comparison is summarized in the main panels of Fig.~\ref{fig:comp_nI}, where the $x$-axis labeled by 
$U$ denotes the strength of the interaction after the quench.
The steady-state and thermal kinetic energy show a leading $U^2$-dependence and within the accuracy of 
our numerics the two prefactors agree. This second order thermalization is in full agreement 
with our  general considerations. 
This can be understood in more detail considering the spin-spin 
correlations. The steady-state as well as thermal expectation values of the three $r$ shown display 
a linear $U$-dependence. Within the accuracy of the results the prefactors of the steady-state and thermal 
values agree; the $C_r^{zz}$ thermalize to linear order in accordance with Sect.~\ref{sec:firstordertherm}. 
For the Hamiltonian (\ref{HIFnNSpin}) $C_1^{zz}$ is directly linked to $V$ of 
Eq.~(\ref{gVdef}). Thus  Eq.~(\ref{21expl}) holds and thermalization of $\tilde h_{\rm kin}$ to second order 
can be concluded from our considerations of Sect.~\ref{sec:secondordertherm}. 

The numerical findings for the XXZ-chain are thus in full agreement with our 
`small-quench thermalization phenomenology'. A more sophisticated analysis to extract 
and compare the differences in the next to leading order (as performed in Sect.~\ref{sec:tightbinding}) 
is currently beyond the accuracy of the DMRG results, as very small but finite contributions 
from the transient dynamics spoil this extremely sensitive numerical test.

\begin{figure}[t]
\centering
\includegraphics[width=\columnwidth]{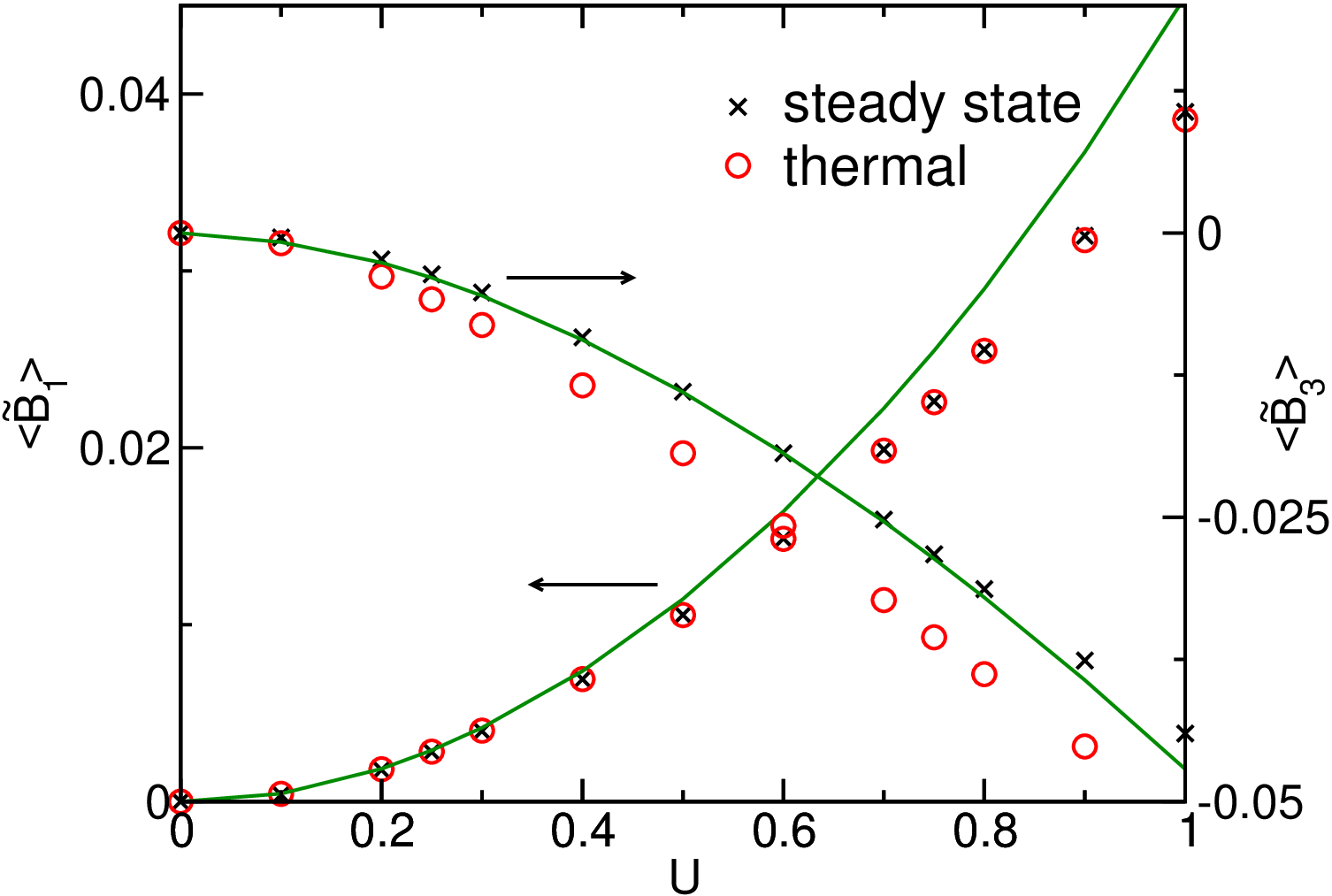}
\caption{The same as in Fig.~\ref{fig:comp_nI} but for the `factor of two class' operators $\tilde B_r$
with odd $r$ (the expectation values for even $r$ vanish by symmetry). 
The solid line shows twice the ground state (with respect to $H_{\rm f}$) expectation value
[see Eq.~(\ref{moeckelallg}) with with $\overline{\left< \ldots \right>}$ 
replaced by $\left< \ldots \right>_{\rm ss}$].
The operator  $\tilde B_1$  is directly related to the kinetic energy and thus in addition 
from the `thermalization class'. The prefactor of the leading $U^2$-terms of the steady-state and 
thermal expectation values agree (crosses and circles coincide for sufficiently small $U$). This does not 
hold for  $\tilde B_3$. }
\label{fig:fac2_comp_nI}
\end{figure}

In Fig.~\ref{fig:comp_nI} we consider $U$ being as large 
as $4$, which means that the quench leads out of the gapless into the gapped phase of the system. We 
do so to illustrate how deviations of the thermal and steady state expectation values continuously 
develop at larger values of $U$.
Surprisingly, the quantitative agreement between steady state and thermal expectation values of 
the local observables $\tilde h_{\rm kin}(t)$ and $C_r^{zz}$ with $r\in \{1,2\}$ is very good 
over a rather large range of $U$. The differences in the prefactors of the higher order terms -- 
$\mathcal{O}(U^3)$ and $\mathcal{O}(U^2)$ for 
$\tilde h_{\rm kin}(t)$ and $C_r^{zz}$ with $r\in \{1,2\}$, respectively -- must be small. In fact, the 
expectation values almost perfectly agree in the entire gapless phase $U<1$. 
For $C_r^{zz}$, 
thermalizing to order $U$, we observe that the smaller $r$, that is the more local the observable, 
the longer the linear term prevails and the longer the two expectation values agree. This is in 
full accordance with the results of  Sect.~\ref{sec:sec}. It also illustrates that 
for small $U$ it was 
virtually impossible to reliably distinguish between thermal and nonthermal steady-state 
behavior in earlier purely numerical studies of the metallic XXZ-chain and related models. 

Next, we turn our attention to the bond operators $\tilde B_r$, which as argued above 
are for any $r$ element of the 
`factor of two class' (compare crosses and lines in 
Fig.~\ref{fig:fac2_comp_nI}), while $r=1$ is additionally in the `thermalization class'.
The results are  summarized  for $r=1$ and $r=3$ in Fig.~\ref{fig:fac2_comp_nI}. The data 
is fully consistent with a $U^2$-dependence. For $\tilde B_1$ we expect that the prefactors 
of the thermal and the steady-state expectation values agree. Figure~\ref{fig:fac2_comp_nI}
indicates this. Also the increasing deviation between $\left< \tilde B_3 \right>_{\rm ss}$
and $\left< \tilde B_3 \right>_{\rm th}$ at increasing $U$ is consistent with the absence 
of second order thermalization for $\tilde B_3$ (not a `thermalization class' observable). 
However, $\left< \tilde B_3 \right>_{\rm ss}$ can still be related to the ground state (with respect 
to $H_{\rm f}$) expectation value by Eq.~(\ref{moeckelallg}). Also for  
`factor of two class' observables we again observe the general trend that the more local the 
observable the longer the steady-state and thermal expectation values agree when 
increasing the quench amplitude (see also Sect.~\ref{sec:sec}).  In addition, due to the quadratic
dependence on the quench parameter at small quenches the expectation 
values themselves are small 
(see  Fig.~\ref{fig:fac2_comp_nI}).

\subsection{Steady state for the model including next-nearest-neighbor interaction}
\label{sec:next}

\begin{figure}[t]
\centering
\includegraphics[width=\columnwidth]{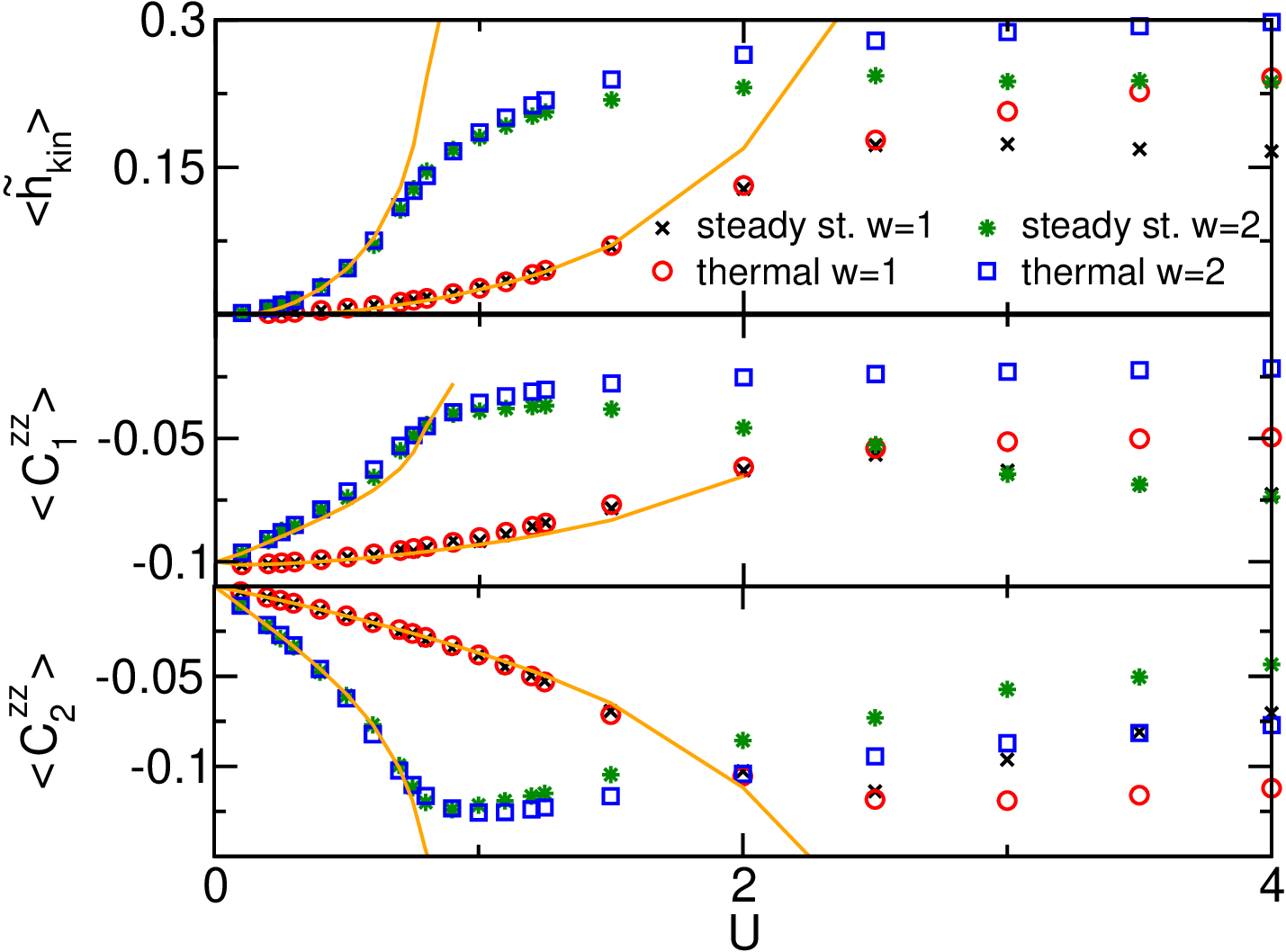}
\caption{The same as in Fig.~\ref{fig:comp_nI} but for the model including a next-nearest-neighbor interaction.}
\label{fig:comp_nnI}
\end{figure}

We perform an analysis similar to the one of the last section in the presence of the next-nearest-neighbor 
interaction $W$. For this model no extensive set of (quasi-) local integrals of motion 
is known. It is generally believed that such a set does not exist. 
We perform the time evolution with $H_{\rm f} = H$ 
of Eq.~(\ref{HIFnnNSpin}) with $U>0$ as well as $W=wU>0$. We show exemplary results 
for the steady-state and thermal expectation values of $\tilde h_{\rm kin}(t)$ and $C_r^{zz}$, with $r=1,2$, 
in Fig.~\ref{fig:comp_nnI}. Again we find that our numerical results for $\tilde h_{\rm kin}(t)$ being 
out of the `thermalization class' (as well as the `factor of two class') are consistent  
with second order thermalization. For spin-spin correlations, not being in the `factor of two class', 
only the leading linear order of $\left< \tilde C_{r}^{zz} \right>_{\rm ss}$ and $\left< \tilde C_{r}^{zz} 
\right>_{\rm th}$ agree. As for $W=0$, the quantitative agreement between thermal and steady-state expectation 
values of  $\tilde h_{\rm kin}(t)$ as well as $C_r^{zz}$ with $r=1,2$ 
are surprisingly good even for interactions as large as $U=2$ ($U=1$) for $W=U$ ($W=2U$), indicating that 
differences in the prefactors of subleading contributions must be small.

The results of this section illustrate that leading order thermalization investigated by us is 
insensitive to the presence or absence of an extensive set of (known) integrals of motion.

\section{Summary}
\label{sec:summary}

A complete summary of our work on small quenches was already given in the introductory 
Sects.~\ref{smallquenchintro} to \ref{classes3_4} and we here refrain from repeating this. 
Instead we emphasize our main results and put them into a broader 
perspective. 

Based on analytical insights for time-averages we have developed a 
`small-quench thermalization phenomenology' which shows that 
{\it leading} (nonvanishing) order thermalization 
of {\it local} observables and correlation functions is a ubiquitous phenomenon for 
quenches out of the ground state of $H_{\rm i}$. This 
holds independently of the number and nature of the integrals of motion inherent to 
a certain model as was shown by explicitly studying the quench dynamics of four distinct 
1d models from different classes. 

In accordance with the prethermalization conjecture the 
steady-state expectation value is reached in a single-step relaxation 
procedure and is thus unaffected by a possible second step of the dynamics 
of the nonlocal mode occupancies (which we did not investigate). 

The more 
local the observable the smaller either the differences between the steady-state 
and thermal expectation values or the smaller the values themselves even for sizable 
amplitudes $g$ of the quench parameter. Both makes it very difficult to make 
reliable statements about thermalization for small $|g|$ based on purely computational 
studies (prone to numerical errors) of models which cannot be solved exactly. The 
heavily studied interaction quench in the XXZ-chain and related models is a particularly 
astonishing example for this as the steady-state and thermal expectation values of 
typical local observables are virtually indistinguishable in the entire metallic regime. 

We start the time evolution in a pure state and the unitary dynamics implies a pure 
state also at large times. Thermalization of local observables can still be expected 
as the rest of the system can be viewed as a reservoir to the (small) subsystem which 
supports the observable. This argument cannot be used for the nonlocal mode 
occupancies and other nonlocal observables. We thus believe that our results 
for small quenches provide a rather satisfying picture of thermalization.   

The expectation values of the 
observables computed in many of our model studies can be obtained from the 
statistical average taken with a properly chosen GGE. This is obvious for 
the quadratic as well as the effectively quadratic models (and mentioned in Sects. 
\ref{sec:tightbinding} and \ref{sec:TLmodel})\cite{Barthel08,Sotiriadis14} 
and, most likely, also holds true for the XXZ-chain with nearest-neighbor 
interaction.\cite{Pozsgay14,Wouters14}. This raises the  
question under what conditions (for which observables) a GGE prediction happens 
to be identical to a thermal description up to a given order in the quench amplitude.  
It would be very interesting to further pursue this question.

We finally emphasize that our explicit results of second order thermalization
for the studied observables of 1d Fermi systems of Sects.~\ref{sec:TLmodel} 
and \ref{sec:lattice} are not at odds with earlier indications of (nonthermal) 
Luttinger liquid universality of the same observables in the steady state after 
an interaction quench.\cite{Karrasch12,Kennes13} The  Luttinger liquid universality
can be found in the large distance behavior of post-quench correlation functions
while the thermal behavior is restricted to local correlation functions and 
observables.

\begin{acknowledgements}
This work was supported by the Deutsche Forschungsgemeinschaft via RTG 1995 (D.K. and V.M.), 
KE 2115/1-1 (D.K.), the Emmy-Noether program, and KA 3360/2-1 (C.K.). We thank S. Kehrein 
and K. Sch\"onhammer for discussions. Simulations were performed with computing resources 
granted by RWTH Aachen University under project rwth0013 and rwth0057. 
\end{acknowledgements}

{}

\end{document}